\documentclass[a4paper,11pt]{article}
\pdfoutput=1 % to allow compilation in JCAP
\usepackage{jcappub}

\usepackage{amsmath}
\usepackage{bm}
\usepackage{graphicx}
\usepackage{enumitem}
\usepackage{geometry}
\usepackage{xspace}
\usepackage{pdflscape}
\usepackage{placeins}
\usepackage{epstopdf}
\usepackage{multicol}
\usepackage{multirow}
\usepackage{mathtools}

\newcommand{\bal}{\begin{aligned}}
\newcommand{\eal}{\end{aligned}}
\newcommand{\F}{\mathcal{F}}

\makeatletter
\gdef\@fpheader{}
\g@addto@macro\bfseries{\boldmath}
\makeatother

\let\oldsqrt\sqrt
\def\sqrt{\mathpalette\DHLhksqrt}
\def\DHLhksqrt#1#2{%
\setbox0=\hbox{$#1\oldsqrt{#2\,}$}\dimen0=\ht0
\advance\dimen0-0.2\ht0
\setbox2=\hbox{\vrule height\ht0 depth -\dimen0}%
{\box0\lower0.4pt\box2}}

\newcommand{\sss}[1]{{\scriptscriptstyle{#1}}}

\newcommand{\uPl}{\mathrm{Pl}}

\newcommand{\usssPl}{\sss{\uPl}}

\newcommand{\Mp}{M_\usssPl}

\newcommand{\beq}{\begin{eqnarray}}
\newcommand{\eeq}{\end{eqnarray}}
\newcommand{\bea}{\begin{equation}\begin{aligned}}
\newcommand{\eea}{\end{aligned}\end{equation}}

\newlength{\wsingfig}
\newlength{\wdblefig}
\newlength{\wquadfig}
\newlength{\wtriplefig}
\subheader{}

\title{Multifield inflation beyond $N_\mathrm{field}=2$: non-Gaussianities and single-field effective theory}

\author[a]{Lucas Pinol}

\affiliation[a]{Institut d'Astrophysique de Paris, UMR 7095-CNRS,
Sorbonne Universit\'e,
 98bis boulevard Arago, 75014
  Paris, France}

\emailAdd{lucas.pinol@iap.fr}

\date{today}

\begin{document}

\sloppy

\abstract{In this article, we study in detail the linear dynamics and cubic interactions for any number $N_\mathrm{field}$ of scalar fields during inflation, directly in terms of the observable curvature perturbation $\zeta$ and $N_\mathrm{field}-1$ entropic fluctuations, a choice that is more suitable for analytical works.
In the linear equations of motion for the perturbations, we uncover rich geometrical effects beyond terms involving just the scalar curvature of the field space, and that come from the non-canonical kinetic structure of the scalar fields when the dimension of the field space is larger than two.
Moreover, we show that a fast rotation of the local entropic basis can result in negative eigenvalues for the entropic mass matrix, potentially destabilising the background dynamics when  $N_\mathrm{field} \geqslant 3$.
We also explain how to render manifest the sizes of cubic interactions between the adiabatic and the entropic fluctuations, extending a previous work of ours to any number of interacting fields.
As a first analytical application of our generic formalism, we derive the effective single-field theory for perturbations up to cubic order when all entropic fluctuations are heavy enough to be integrated out.
In a slow-varying limit, we recover the cubic action expected from the effective field theory of inflation, but with a prediction for the usual Wilson coefficients in terms of the multifield parameters, thus proposing a new interpretation of the bispectrum in this generic $N_\mathrm{field}$ context.
}


% \arxivnumber{XXXX.XXXXX}

\maketitle

\section{Introduction}

Inflation, an era of almost-exponential expansion of spacetime, constitutes our paradigm for the early universe.
Not only does it explain the observed spatial flatness and large-scale homogeneity of the universe at the time of recombination and photon decoupling, but it also predicts with a high accuracy the Gaussian statistics of the tiny anisotropies in the Cosmic Microwave Background (see~\cite{Akrami:2018odb,Akrami:2019izv} for recent observational constraints).
Despite these successes, the exact mechanism behind inflation is yet to be understood.
Indeed, the simplest class of models, the framework of so-called single-field slow-roll inflation with canonical kinetic terms, suffers from a number of theoretical limitations as it relies on the existence of a unique degree of freedom in the early universe, the inflaton, which must have a very flat potential.
First, realistic UV completions of inflation that are needed to describe phenomena at higher energies and from which inflation must be inferred (top-down approach), typically predict the presence of not only one, but several scalar fields, and they often come with non-canonical kinetic terms (for a review see, \textit{e.g.},~\cite{Baumann:2014nda}).
Secondly, even just as a low-energy Effective Field Theory (EFT), the single-field slow-roll picture (bottom-up approach) is not satisfactory neither as Planck-suppressed operators renormalize the potential in a way that it can not remain flat in Planck mass units (see~\cite{Copeland:1994vg,Stewart:1994ts} for the historical papers and \cite{McAllister:2007bg,Baumann:2009ni} for pedagogical reviews on the so-called $\eta$-problem).
In this regard, understanding the physics of inflation provides us with a formidable opportunity to test and investigate physics Beyond the Standard Model.

Amongst the possible ways to evade the aforementioned theoretical limitations, and motivated by top-down constructions, we shall consider here multifield models of inflation with non-canonical kinetic terms, \textit{i.e.} with a curved field space.
Multifield inflationary dynamics present striking characteristic observational signatures, such as correlated adiabatic and isocurvature perturbations, features in the primordial power spectrum and order one non-Gaussianities with specific patterns (see, \textit{e.g.}, \cite{Langlois:1999dw,Wands:2002bn, Lesgourgues:1999uc,Cremonini:2010sv,Peterson:2010np,Achucarro:2010da, McAllister:2012am, Chen:2009we,Chen:2009zp,Arkani-Hamed:2015bza,Flauger:2016idt,Lee:2016vti,Arkani-Hamed:2018kmz}).
More exotic features can also be found, such as a large flattened shape for the bispectrum and higher-order correlation functions~\cite{Garcia-Saenz:2018ifx,Garcia-Saenz:2018vqf,Fumagalli:2019noh,Ferreira:2020qkf}, a production of Primordial Black Holes (PBH)~\cite{Palma:2020ejf,Fumagalli:2020adf} and of small-scale Primordial Gravitational Waves (PGW)~\cite{Braglia:2020eai}.
Actually, most multifield studies focus on the simplest non-trivial setup with $N_\mathrm{field}=2$ scalar fields, and a few look at $\mathcal{O}(N_\mathrm{field})$ models in the $N_\mathrm{field}\rightarrow \infty$ limit (see, \textit{e.g.},~\cite{Choi:2008et, Easther:2013rva, Christodoulidis:2019hhq}), whereas here we will work in full generality with $N_\mathrm{field}$ being any positive integer.
Of course, this generic setup has already been investigated, with the computation of the second-order and third-order actions for multifield inflation in the flat gauge~\cite{Sasaki:1995aw,GrootNibbelink:2000vx, GrootNibbelink:2001qt,Gong:2011uw,Gao:2012uq,Elliston:2012ab}, and even used in the numerical implementation of the transport approach~\cite{Dias:2016rjq,Seery:2016lko,Mulryne:2016mzv,Ronayne:2017qzn,Butchers:2018hds}.
However, the comoving gauge in which the observable adiabatic perturbation $\zeta$ naturally appears is best suited for analytical works, such as Quasi-Single-Field (QSF) calculations of the adiabatic power spectrum and bispectrum~\cite{Chen:2009we,Chen:2009zp,Noumi:2012vr}, or integrating out heavy fluctuations~\cite{Achucarro:2012sm, Burgess:2012dz, Gwyn:2012mw, Cespedes:2013rda, Garcia-Saenz:2019njm} (see~\cite{Pi:2012gf, Gong:2013sma} for consistency checks between the two approaches in a regime of common validity, and~\cite{Tolley:2009fg} for integrating out the whole extra heavy field and not just its fluctuations).
In our previous article~\cite{Garcia-Saenz:2019njm}, we revisited in detail the bispectrum produced in generic models of inflation with $N_\mathrm{field}=2$ scalar fields, by computing the cubic action in the comoving gauge.
In particular, we unveiled the effect of the Ricci scalar of the field space on the inflationary bispectrum.

\vspace{0.2cm}
In the present article, we combine and extend previous studies by studying $N_\mathrm{field}$-inflation with curved field space, directly in the comoving gauge and in terms of $\zeta$ and $N_\mathrm{field}-1$ entropic perturbations, highlighting the special features due to the dimension of the field space.
We begin in Sec.~\ref{sec2} by reviewing some notations as well as the background dynamics of multifield inflation.
We explain the decomposition along adiabatic and entropic directions and define the rate of turn of the local entropic basis with the matrix $\Omega$.
 Covariant perturbations beyond linear order are carefully chosen, and the ADM formalism applied to cosmology is reviewed with particular attention afforded to the boundary terms.
Next, we decompose perturbations in the comoving gauge and along the unit vectors of the local orthonormal basis, identifying our fundamental degrees of freedom; $\zeta$ and $\F^\alpha$ for $\alpha \in \{1,...,N_\mathrm{field}-1\}$.
We present the quadratic Lagrangian, Eq.~\eqref{L2}, which incorporates the mixing of the massless adiabatic perturbation with the entropic sector, as well as self-mixings of the massive entropic perturbations, together with the corresponding equations of motion (EoM) in Eqs.~\eqref{linear-eom-zeta}-\eqref{linear-eom-F}.
We pinpoint the role of the field-space geometry beyond its scalar curvature whenever $N_\mathrm{field}\geqslant 3$ and comment about a conceivable destabilisation of background trajectories due to a fast rotation of the entropic basis, a possibility that was overlooked so far to our knowledge.
Lastly, we derive for completeness the corresponding quadratic Hamiltonian that we use to define our free theory.

Sec.~\ref{sec3} is devoted to the computation of the cubic action in the comoving gauge, for our general setup of $N_\mathrm{field}$ scalar fields with non-canonical kinetic terms, without assuming special features for the background dynamics such as slow-varying approximations.
This corresponds essentially to the $N_\mathrm{field}$-generalisation of the famous Maldacena's calculation of the bispectrum in canonical single-field inflation~\cite{Maldacena:2002vr} and of our previous work in the two-field case~\cite{Garcia-Saenz:2019njm}.
We explain that the brute-force expansion of the action up to cubic order is not sufficient to unveil the true sizes of the interactions, and show how to render them manifest with uses of integrations by parts and of the linear EoM, carefully keeping the time-boundary terms that are needed for a full computation of the bispectrum.
The final result~\eqref{L3-final}-\eqref{boundary-final} is one of the main achievements of this article: it presents the genuine sizes of cubic interactions between adiabatic and entropic perturbations without any approximation.

This form of the cubic action is particularly suited for analytical works, and we show in Sec.~\ref{sec4} a first application of it.
There, we derive the effectively single-field theory for fluctuations that is found by integrating out heavy entropic perturbations, a procedure that is justified when the entropic mass matrix dominates over the other contributions in the EoM for $\F^\alpha$.
We comment on the more precise conditions of validity for this single-field regime, and infer a reduced speed of sound, $c_s^2<1$, for the free propagation of $\zeta$ whenever the background trajectory is bending.
At the level of cubic interactions, the single-field effective theory matches the one of a truly single degree of freedom $\phi$ with a $P(X,\phi)$ Lagrangian~\cite{ArmendarizPicon:1999rj,Garriga:1999vw,Seery:2005wm,Chen:2006nt}, provided some matching between the multifield setup and the derivatives of the function $P$.
In a slow-varying approximation, we recover the leading-order operators expected from the EFT of inflation~\cite{Creminelli:2006xe,Cheung:2007st}, with a prediction for the Wilson coefficient $A$ in terms of the multifield potential and the field-space geometry.

Sec.~\ref{sec5} is devoted to conclusive remarks and an outlook.

\section{$N_\mathrm{field}$-inflation in curved field space}
\label{sec2}

%Understanding the particle content of the early universe, "quasi-single field inflation", or "cosmological collider physcis". Previous studies focus on one extra massive scalar field. Squeezed limit of the bispectrum. Either only one kind of interaction is considered (Wang and Chen), or all interactions are considered in an EFT approach without physical interpretation for their origins (the 3 Japanese). Here we not only consider any number $N$ of scalar fields, but we explicitely derive their interactions in an adiabatic-entropic decomposition, as functions of the UV parameters of multifield inflation with curved field space. In particular, this setup provides a natural interpretation of the mixing couplings as covariant rates of turn of the local orthonormal basis projecting along adiabatic and entropic directions.

In this section, we first review standard notations for multifield inflationary setups and choose a decomposition of the background fields and perturbations in a local orthornormal basis consisting of an adiabatic direction (instantaneously following the background trajectory) and $N_\mathrm{field}-1$ entropic (perpendicular) directions.
Field-space covariant and perturbatively spacetime-gauge-invariant fluctuations are identified, and constraint equations needed to remove non-dynamical degrees of freedom,  are shown.
We comment on the necessity of the Gibbons-Hawking-York boundary term in the action, in order to have a well-defined variational principle.
A first interesting point in this generic $N_\mathrm{field}$ setup is the quadratic action displayed in Eqs.~\eqref{L2}-\eqref{Omega2} and that matches the only explicit reference in the literature after a few changes of variables~\cite{Achucarro:2018ngj}.
We outline geometrical effects beyond the scalar curvature of the field space depending on the number of fields, and identify potential background instabilities from the large-scale limit of the linear equations of motion~\eqref{linear-eom-zeta}-\eqref{linear-eom-F} that we found.
Eventually, we show the corresponding quadratic Hamiltonian that we use to define our free theory for this work, and that could be used in future works to compute analytically the power spectrum in a perturbative scheme in terms of the mixing parameters $\omega_\alpha$ with $\alpha \in \{1,..., N_\mathrm{field}-1\}$, or numerically with the transport approach.

\subsection{Generalities}

\paragraph{Setup.}

In this paper, we consider general multifield non-linear sigma models of inflation, described by the action
\beq
S=\int {\rm d}^4x\sqrt{-g}\bigg[\frac{\Mp^2}{2}\,R(g)-\frac{1}{2}\,G_{IJ}(\phi)\nabla^{\mu}\phi^I\nabla_{\mu}\phi^J-V(\phi)\bigg]\,,
\label{S}
\eeq
with $G_{IJ}$ the metric of the internal field space manifold. The latin index $I$ runs from 1 to $N_\mathrm{field}$ the total number of scalars.
Our convention for the Riemann tensor is
\beq
R^I_{\phantom{I}JKL}=\Gamma^I_{JL,K}+\Gamma^I_{KM}\Gamma^M_{JL}-(K\leftrightarrow L)\,,
\eeq
where we denote by $\Gamma^I_{JK}$ the corresponding Levi-Civita connection. We also use the summation convention for repeated indices.

\paragraph{Background.}

The space-time background is assumed to be of the Friedmann-Lema\^itre-Robertson-Walker (FLRW) type, with a scale factor $a(t)$ and related Hubble parameter $H(t)=\dot{a}(t)/a(t)$, a dot representing a derivative with respect to cosmic time.
This background and its dynamics are ruled by the presence of the homogeneous parts of the $N_\mathrm{field}$ scalar fields, $\bar{\phi}^I(t)$.
Their evolution is given by 
\beq \label{back-eom-fieldframe}
\mathcal{D}_t \dot{\bar{\phi}}^I+3H\dot{\bar{\phi}}^I + G^{IJ}V_{,J}=0
\,,
\eeq
where the time field-space covariant derivative action on a field-space vector $A^I$ is given by $\mathcal{D}_t A^I = \dot{A}^I +\Gamma^I_{JK}\dot{\bar{\phi}}^J A^K$, and where a comma denotes a field-space derivative: $V_{,J}=\partial V / \partial \phi^J$.
The Friedmann equations can be written as
\beq \label{back-Friedmann}
\dot{\sigma}^2=2\Mp^2H^2\epsilon\,,\qquad V=M_P^2H^2(3-\epsilon)\,,
\eeq
where we have introduced the first Hubble slow-roll parameter, $\epsilon=-\dot{H}/H^2$ as well as the total background field velocity $\dot{ \sigma}=\sqrt{G_{IJ}\dot{\bar{\phi}}^I\dot{\bar{\phi}}^J}$.
Note that although what one can have in mind is a sufficiently long period of inflation for which both $\epsilon$ and $\eta=\dot{\epsilon}/(H \epsilon)$ must be small, all the results presented in this paper are exact in the sense that they are not expansions in those parameters, except if explicitly mentioned.
In particular, our results are perfectly applicable to the description of transient departures from slow-roll, other attractors such as ultra-slow-roll, or even multi-scalar dynamics beyond inflation (see~\cite{Hamada:2020phg} for a recent application of our previous work~\cite{Garcia-Saenz:2019njm} in the case of $N_\mathrm{field}=2$ to holographic theories).

\paragraph{Adiabatic-entropic decomposition.}

\begin{figure}
\hspace{-0.65cm}
\includegraphics[scale=0.45]{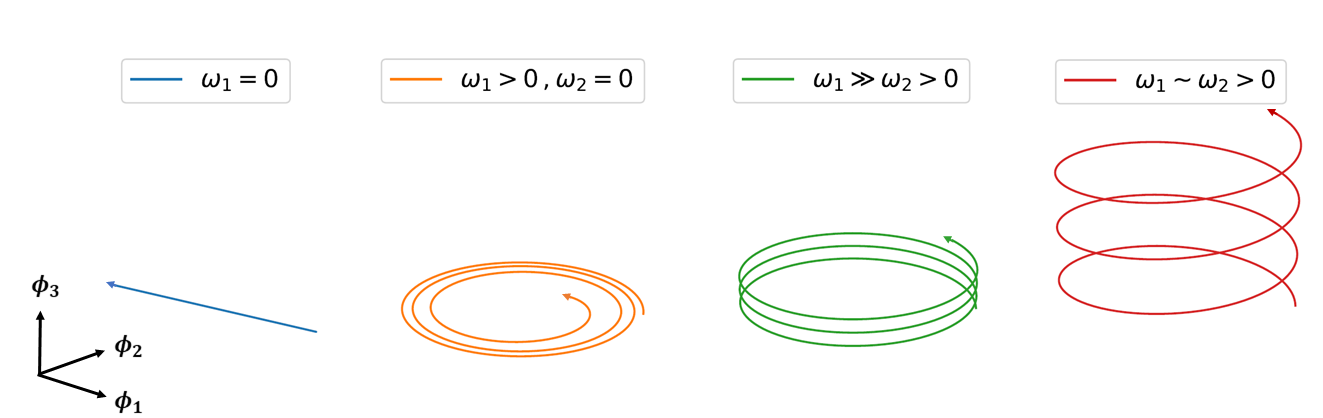}
\caption{Examples of trajectories in a flat field-space manifold with $N_\mathrm{field}=3$, and for different values of the curvature $\omega_1$ and of the torsion $\omega_2$.}
\label{fig:basis}
\end{figure}

It is convenient to introduce vielbeins along the background trajectory to define a local orthonormal basis: $e^I_a=\left(e_\sigma^I,e_{s_\alpha}^I\right)$ with $\sigma$ denoting the adiabatic direction  and $s_\alpha=s_1,...,s_{N_\mathrm{field}-1}$ the $N_\mathrm{field}-1$ entropic directions (that we shall denote $s_\alpha \rightarrow \alpha$ for simplicity in the following).
One can decompose the field-space metric as $G^{IJ}=\delta^{ab}e^I_a e^J_b=e^I_\sigma e^J_\sigma +e^I_\alpha e^J_\alpha$.
Note that in this kind of expressions there is a sum over the $N_\mathrm{field}-1$ values for $\alpha$ but that $\sigma$ denotes only one direction: the adiabatic one.
Of course, such local orthonormal basis is not unique and any orthonormal rotation conserves its properties.
In particular, although the adiabatic direction has the physical intepretation of pointing towards the instantaneous direction of the background trajectory, there is no preferred frame for the entropic sector.
Therefore we fix this ambiguity by defining the first entropic direction as the covariant rate of turn of the adiabatic direction (that vanishes if the background trajectory coincides with a geodesic of the field space), and all other vectors can be defined as the wedge product of the two previous ones\footnote{Note that, strictly speaking, when the bending $\omega_1$ of the background trajectory vanishes exactly, the individual entropic directions can not be defined in an unambiguous way. However this has no practical implication as in that fine-tuned case the evolution of $\zeta$, the observable curvature perturbation, is linearly decoupled from the dynamics of the entropic sector as we shall find soon.} (see~\cite{Kaiser:2012ak, Achucarro:2018ngj} for similar constructions in the inflationary context).
This results in a particular form for the covariant rate of turn of the local orthonormal basis:
\begin{equation}
\label{cov-rate-matrix}
\mathcal{D}_t \left(
\begin{array}{cc}
   e_\sigma^I \\ \hline  \\ [-10pt]
   e_1^I \\
   e_2^I \\
   \vdots & \\
   e^I_{N_\mathrm{field}-1} &
\end{array}
\right) 
=
\left(
\begin{array}{c|cccc}
   0 & \omega_1 & 0 & \hdots & 0 \\  \hline \\ [-10pt]
   - \omega_1 & 0 & \omega_2 & \hdots &0 \\
   0 & - \omega_2 & 0 & \hdots & 0 \\
   \vdots &  & \ddots &\ddots & \vdots \\
   0 & 0 & \hdots & - \omega_{N_\mathrm{field}-1} & 0 \\
\end{array}
\right) \left(
\begin{array}{cc}
   e_\sigma^I \\ \hline \\ [-10pt]
   e_1^I \\
   e_2^I \\
   \vdots & \\
   e^I_{N_\mathrm{field}-1} &
\end{array}
\right) 
\,,
\end{equation}
where the turning rates $\omega_\alpha$ have the dimension of a mass. One clearly sees that the first entropic direction plays a special role because it bridges together the adiabatic and purely entropic sectors. It will prove useful to use the coordinate expression of this matrix equation, separating the adiabatic direction from the entropic sector:
\begin{align}
\mathcal{D}_t e_\sigma^I&=\omega_1 e_1^I \\
\mathcal{D}_t e_{\alpha}^I&=- \delta_{1\alpha} \omega_1 e_\sigma^I + \Omega_\alpha{}^\beta e_\beta^I \nonumber
\,,
\end{align}
$\Omega$ being the anti-symmetric square matrix of size $(N_\mathrm{field}-1) \times (N_\mathrm{field}-1)$ in the lower right corner of the rotation matrix in \eqref{cov-rate-matrix}.
This decomposition is equivalent to a Cartan frame decomposition and one can interpret each of the $N_\mathrm{field}-1$ mixing parameters $\omega_\alpha$ as the turning rates of the local basis.
In mathematical studies of parametric curves in $N_\mathrm{field}$ dimensions, the parameters $\omega_\alpha$ are called ``generalized curvatures" of the trajectory (in the special case $N_\mathrm{field}=3$, $\omega_1$ is called the curvature and $\omega_2$ the torsion, see Fig.~\ref{fig:basis} for a geometrical interpretation of these parameters), but here we will name them ``bendings" or ``mixing parameters" not to be confused with statements about the geometry of the curved field-space manifold that are independent of a given trajectory.
Note also that by definition the local metric in the adiabatic-entropic basis is flat, $G_{IJ}e^I_a e^J_b= \delta_{ab}$, this implying in particular that ``entropic indices" $\alpha$ can be up or down for convenience without making any difference.

It is then possible to decompose the EoM of inflation along those directions. Using that by definition $\dot{\bar{\phi}}^I=e_\sigma^I \dot{\sigma}$, one can rewrite the background equations of motion for the scalar fields~\eqref{back-eom-fieldframe} as
\beq
\label{back-eom-localframe}
\ddot{\sigma}+3H\dot{\sigma} + V_{,\sigma}=0  \text{  ,  } \hspace{1.5cm} \sqrt{2\epsilon}H \Mp \omega_1 \delta_{\alpha,1} + V_{,\alpha} = 0 \,.
\eeq
where we defined projections of the potential gradients along the adiabatic and entropic directions, $V_{,\sigma}=e^I_\sigma V_{,I}$ and $V_{,\alpha}=e^I_\alpha V_{,I}$. Note in particular that this last quantity is non-zero only when projected along the first entropic direction, according to~\eqref{back-eom-localframe}. This is a convenient feature from our choice for the entropic frame.

\subsection{Covariant, gauge-invariant perturbations}

\paragraph{Covariance.} When going beyond the study of linear fluctuations, one should be careful to use variables that are covariant under field-space redefinitions.
Although not a requirement \textit{per se}, as predictions for observable quantities do not depend on particular choices of variables, it is useful and conceptually clearer to deal with covariant objects.
The concern and its resolution have been first described in Ref.~\cite{Gong:2011uw} for generic multifield models.
In any given gauge, the idea is to use, not the field fluctuations $\delta\phi^I=\phi^I-\bar{\phi}^I$, which do not transform covariantly beyond linear order, but the vectors $Q^I$ living in the tangent space at $\bar{\phi}^I$ and that correspond to the ``initial velocity" of the geodesic connecting the two points labelled by $\bar{\phi}^I$ and $\phi^I$ (this geodesic is unique if the separation between the two points is sufficiently small).
Up to third order in fluctuations, one finds:
\beq
\delta\phi^I=Q^I-\frac{1}{2}\,\Gamma^I_{JK}Q^JQ^K+\frac{1}{6}(2\Gamma^I_{LM}\Gamma^M_{JK}-\Gamma^I_{JK,L})Q^JQ^KQ^L+{\cal O}(Q^4)\,.
\label{deltaphi-cubic}
\eeq 
Note that although $Q$ is sometimes used to denote inflationary fluctuations in the flat gauge, here they denote non-linearly covariant inflatons' fluctuations in any gauge. We shall return soon to this matter of gauge choice.

\paragraph{ADM formalism and genuine degrees of freedom.}
Let us quickly review the ADM formalism applied to the cosmological context, paying particular attention to the boundary terms.
Regular cosmological spacetimes' manifolds $\mathcal{M}$ equipped with a metric $g$ can be foliated by Cauchy surfaces $\Sigma_t$ parameterized by a global time function $t$.
One can then define the unit normal vector $n^\mu$ to the hypersurface $\Sigma_t$, and an induced spatial 3-metric $h_{ij}$ on each of those hypersurfaces by $h_{ij}=g_{ij}+n_i n_j$, as well as a lapse $N$ and shift $N^i$ such that the spacetime metric takes the following form~\cite{adm,Salopek:1990jq}:
\beq
{\rm d}s^2=-N^2 {\rm d}t^2+h_{ij}({\rm d}x^i+N^i {\rm d} t)({\rm d}x^j+N^j {\rm d}t)\,.
\eeq
The idea of the ADM formalism is to interpret the spatial $3$-metric $h_{ij}$ as the dynamical quantity that evolves on a manifold $\Sigma$ when moving forward in time, while the lapse $N$ and the shift $N^i$ only describe how to move forward in time.
This intuition shall be confirmed soon by the appearance of the constraint equations that enable one to express the lapse and the shift as linear combinations of the true dynamical perturbations.
Then, it is possible to decompose the $4$-dimensional Ricci scalar $R$
%$=2\left(G_{\mu\nu} n^\mu n^\nu + R_{\mu\nu}n^\mu n^\nu\right)$ 
according to this foliation using the Gauss-Codazzi relation (see, \textit{e.g.},~\cite{Wald}):
\beq\bal
\label{gauss-codazzi}
R&= 2 \left( G_{\mu\nu}- R_{\mu\nu} \right)n^\mu n^\nu \, \text{ with } \\
G_{\mu\nu} n^\mu n^\nu&=\frac{1}{2}\left( R^{(3)} + K^2 - K_{ij}K^{ij}  \right)  \,, \\
 R_{\mu\nu}n^\mu n^\nu&=K^2 - K_{ij}K^{ij} - \nabla_\mu \left(n^\mu \nabla_\nu n^\nu \right) + \nabla_\mu\left( n^\nu \nabla_\nu n^\mu\right) \,,
\eal \eeq
where $K_{ij}$ denotes the extrinsic curvature of the spacelike hypersurfaces $\Sigma_t$, and reads $K_{ij}=\frac{1}{2N} \left( \dot{h}_{ij}-2 N_{(i|j)}\right)$, with the symbol $|$ used for the spatial covariant derivative associated with the $3$-metric $h_{ij}$ and parentheses denoting symmetrization: $N_{(i|j)}=\left(N_{i|j}+N_{j|i}\right)/2$, and with $K=h^{ij}K_{ij}$ its trace.

Usually, one does not care about the total derivatives in the third line of~\eqref{gauss-codazzi}, although they do not vanish in general (note that they come from the derivatives of the Christoffel symbols in the Riemann tensor),
because boundary terms in the action do not affect equations of motion.
But in cosmology in particular, the equations of motion do not contain all the relevant information to derive observational predictions; to compute the meaningful equal-time correlation functions one needs to use the ``in-in" formalism that requires the knowledge of the whole Hamiltonian and not just its bulk terms. More precisely, only the total-time derivatives are important in cosmology since one is not provided with any boundary in the spatial directions, and one can check that the second of the total-derivative terms in the third line of~\eqref{gauss-codazzi} vanishes in the cosmological context since it is orthogonal to the spacelike hypersurfaces $\Sigma_t$.
However the first one remains, and it can be rewritten in terms of the scalar extrinsic curvature which indeed reads  $K=\nabla_\nu n^\nu$ at this boundary.
Eventually, the resulting action can be rewritten:
\beq\bal
S&=S_{\mathrm{EH}}+S_{\mathrm{matter}}  \text{  with   }  \\
S_{\mathrm{EH}}&= \int_\mathcal{M} {\rm d}t {\rm d}^3 \vec{x} \left[ \frac{1}{2} \sqrt{h}N\,\left( R^{(3)} + (K_{ij} K^{ij}-K^2)\right) -\partial_t \left(2 \sqrt{h} N K \right) \right] \\
S_{\mathrm{matter}} &= \frac{1}{2}\int_\mathcal{M} {\rm d}t {\rm d}^3 \vec{x} \sqrt{h}N\,\left(\frac{1}{N^2}G_{IJ}v^Iv^J -G_{IJ}h^{ij} \partial_i \phi^I \partial_j \phi^J-2V \right)  \,,
\label{action-ADM}
\eal
\eeq
where $h=$ det$(h_{ij})$,  $R^{(3)}$ is the Ricci curvature of the spatial hypersurfaces calculated with $h_{ij}$ and $v^I=\dot \phi^I -N^j \partial_j \phi^I$.
The total time-derivative in the Einstein-Hilbert action written in the ADM formalism is known to be potentially problematic because it contains second-order time derivatives of the spatial metric, which could wrongly point towards new spurious degrees of freedom.
Interestingly, this term is exactly canceled if the bulk Einstein-Hilbert action is supplemented by the Gibbons-Hawking-York boundary term $S_{\mathrm{GHY}}=2 \int_{\partial\mathcal{M}} \mathrm{d} \mathcal{V} K $ with appropriate measure $\mathrm{d} \mathcal{V}$ on the spatial boundary $\partial\mathcal{M}$, which is known to be necessary to have a well-defined variational principle for general relativity~\cite{York:1972sj,REGGE1974286,Wald}, or to define the notion of black hole entropy and for the path integral formulation of quantum gravity~\cite{Gibbons:1976ue,Hawking:1995fd}.
%This boundary term has been widely discussed in the general relativity literature~\cite{?,?,?} as incorporating it is needed to correctly define the notions of mass and momentum in asymptotically flat spacetimes~\cite{???}.
In the following, we shall consider implicitly that the description of spacetime incorporates this boundary term, and thus that our resulting action, $S=S_{\mathrm{EH}}+S_\mathrm{GHY}+S_{\mathrm{matter}}$, yields a well-defined variational principle with no extra degree of freedom in the metric, and that our starting point for further computations does not contain any relevant boundary term.

Also, the lapse $N$ and the shift $N^i$ appear without time derivatives in the action, and only linearly once expressed in a Hamiltonian language in terms of the right degrees of freedom and their canonically conjugated momenta. This means as announced that they are non-dynamical and can be perturbatively solved from constraints equations in terms of the genuine degrees of freedom: $\delta S / \delta N = 0 = \delta S / \delta N^i$.

\paragraph{Gauge choice.} Throughout the paper, we neglect tensor and vector perturbations, for the usual reason that they are decoupled from the scalar fluctuations at linear order, so they can only contribute to two-point and three-point correlation functions of the latter by loops, such effects being usually suppressed compared to the tree-level interactions that we take here into account.
As is well known in cosmological perturbation theory, there is an ambiguity in identifying quantities defined on the full spacetime manifold $\mathcal{M}$ and on the unperturbed, FLRW one $\bar{\mathcal{M}}$.
This problem is tackled with an explicit choice of scalar functions $T,L$ for the gauge $\xi^\mu=\left(T,\partial^i L\right)$ that relates $\mathcal{M}$ to $\bar{\mathcal{M}}$, which eliminates two spurious degrees of freedom.
While $N$ and $N^i$ contain both a single scalar non-dynamical degree of freedom as we have just seen, the spatial part of the metric $h_{ij}=a^2\left(e^{2\psi}\delta_{ij}+\partial_i\partial_j E \right)$ contains two dynamical ones.
Thus, a popular gauge choice when studying multifield inflation is the spatially flat gauge, $\psi^{\rm flat}=0=E^{\rm flat}$, such that $h^{{\rm flat}}_{ij}=a^2 \delta_{ij}$, and in which all physical degrees of freedom are the field fluctuations $Q^I_\mathrm{flat}$.
This choice is made in many studies (see \textit{e.g.} \cite{Seery:2005gb,Langlois:2008mn,Langlois:2008qf,Tzavara:2011hn,Elliston:2012ab,Tzavara:2013wxa} for general formalisms) and has a number of advantages. However, as we discussed in the introduction, here we use the ``comoving gauge"~\footnote{Strictly speaking, the comoving gauge is defined upon the vanishing of the $0-i$ component of the stress-energy tensor. We note that referring to our gauge as ``comoving" is a slight abuse of terminology, as it was shown that this condition could not be matched in multifield models beyond linear perturbation theory \cite{Rigopoulos:2005xx,Langlois:2006vv,RenauxPetel:2008gi,Lehners:2009ja}.
However, as shown in these works, setting $e^{\sigma}_I Q^I=0$ defines an approximate comoving gauge on super-Hubble scales in expanding universes, so we decided to keep the terminology ``comoving gauge" for simplicity.} that we define by the following conditions: the adiabatic field fluctuation is set to zero, $e^{\sigma}_I Q^I_{\rm com}=0$, and $E_\mathrm{com}=0$, such that the only surviving fluctuations are
\beq
\label{def-F-zeta}
\text{the $N_\mathrm{field}-1$ entropic field perturbations } &e^{\alpha}_I Q^I_\mathrm{com} \equiv \F^\alpha \,, \nonumber \\
\text{the adiabatic curvature perturbation }  &\psi_{\rm com}\equiv \zeta \,,
\eeq
hence the spatial part of the metric reads $h^{{\rm com}}_{ij}=a^2e^{2\zeta}\delta_{ij}$.

To compute the action up to cubic order, it is sufficient to plug back in the action the expressions of the lapse and the shift at linear order only, in terms of the physical degrees of freedom: $N=1+\alpha$ and $N^i=\delta^{ij} \partial_j\theta/a^2$. The linear constraint equations in our gauge read
\beq
\alpha^{(1)}=\frac{\dot{\zeta}}{H}\,,\qquad \theta^{(1)}=-\frac{\zeta}{H}+\chi\,,
\label{solution-constraints}
\eeq
where the function $\chi$, defined by
\beq
\frac{1}{a^2}\,\partial^2\chi=\epsilon\dot{\zeta}+ \sqrt{2\epsilon}  \frac{\omega_1}{\Mp}\,\F^1\,,
\label{def-chi}
\eeq
only depends on the comoving curvature perturbation $\zeta$ and the first entropic field fluctuation $\F^1$.
Actually, $\chi$ is proportional in Fourier space to the linear canonical momentum associated with $\zeta$ as we shall see soon. It also coincides with $-\Psi/H$ where $\Psi$ is the gauge-invariant Bardeen potential, and since $\partial^2 \chi/a^2$ is negligible on large scales, one finds from Eq.~\eqref{def-chi} the super-Hubble feeding of $\zeta$ by the first entropic fluctuation $\F^1$ when the background trajectory bends in the adiabatic direction ($\omega_1 \neq 0$).

\subsection{Quadratic action}

The second order Lagrangian dictating the linear dynamics of the perturbations, after substitution of the constraint for $\alpha^{(1)}$ (contributions from $\theta^{(1)}$ cancel one another), is found to be:
\beq \bal \label{L2}
\mathcal{L}^{(2)}\left(\zeta,\mathcal{F}^\alpha\right) = & a^3 \left\{\epsilon M_p^2\left(\dot{\zeta}^2 - \frac{(\partial \zeta)^2}{a^2} \right) +  2 \sqrt{2\epsilon} \Mp \omega_1 \delta_{\alpha 1}  \mathcal{F}^\alpha \dot{\zeta} \right. \\
& \left. + \frac{1}{2} \left[ \left(\dot{\mathcal{F}}^\alpha\right)^2 - \frac{(\partial \mathcal{F}^\alpha)^2}{a^2} -m_{\alpha\beta}^2 \mathcal{F}^\alpha \mathcal{F}^\beta   + 2 \Omega_{\alpha \beta} \dot{\F}^\alpha \F^\beta  \right] \right\}\,,
\eal \eeq
with \begin{equation}
\label{mass-matrix}
m^2_{\alpha \beta}=V_{;\alpha \beta} + \Omega_\alpha{}^\gamma \Omega_{\gamma\beta} - \delta_{\alpha 1} \delta_{\beta 1} \omega_1^2 + 2 \epsilon H^2 \Mp^2 R_{\alpha \sigma \beta \sigma} \,,
\end{equation} 
the symmetric mass-matrix elements for the entropic sector. An index $\sigma$ represents a contraction with the adiabatic direction and indices $\alpha$ correspond to contractions with entropic directions: $V_{;\alpha \beta}=e^I_\alpha e^J_\beta V_{;IJ}$ with a semicolon denoting a covariant derivative in field space and $R_{\alpha \sigma \beta \sigma}=e^I_\alpha e^J_\sigma e^K_\beta e^L_\sigma R_{IJKL}$. 
Note also that $\Omega_\alpha{}^\gamma \Omega_{\gamma \beta}$ is nothing but the $(\alpha,\beta)$ component of the square of the $(N_\mathrm{field}-1) \times (N_\mathrm{field}-1)$ matrix $\Omega$, which reads
\begin{equation}
\label{Omega2}
\Omega^2=\left(
\begin{array}{ccccc}
    - \omega_2^2 & 0 & \omega_2 \omega_3 & \hdots & 0 \\  
   0 & -\omega_2^2-\omega_3^2 & 0 & \hdots &0 \\
   \omega_2\omega_3 & 0 & -\omega_3^2-\omega_4^2 & \hdots & 0 \\
   \vdots &  & \ddots &\ddots & \vdots \\
   0 & \hdots & \omega_{N_\mathrm{field}-2}\omega_{N_\mathrm{field}-1} & 0 & - \omega_{N_\mathrm{field}-1} ^2 \\
\end{array}
\right) \,,
\end{equation}
and has only negative (or vanishing) eigenvalues.
%in the lower right corner of~\eqref{cov-rate-matrix}.
Some useful notations are $\left(\partial X\right) \left( \partial Y\right)=\partial_i X \partial_i Y$ and  $\partial^2=\partial_i \partial_i$. Let us also introduce formally $\partial^{-2}$ the inverse Laplacian operator: $\partial^{-2}\partial^2=\partial^2\partial^{-2}=1$, which shall be used in the following. The quadratic Lagrangian~\eqref{L2} matches Ref.~\cite{Achucarro:2018ngj} in the more particular context of holographic multifield inflation, once appropriate changes of variables are made.

\paragraph*{Geometrical effects.} Before proceeding, let us understand better the geometrical effects in the mass matrix of entropic modes, $m^2_{\alpha\beta}$. For this purpose, it will prove useful for $N_\mathrm{field} \geqslant 3$ to decompose the Riemann tensor as
%\begin{align}
%\label{Ricci-decomposition}
%R_{IJKL}=&\frac{R_\mathrm{fs}}{N(N-1)}\left(G_{IJ}G_{KL}-G_{IL}G_{JK}\right) \nonumber \\
%&+\frac{1}{N-2}\left(G_{IK} Z_{JL} - G_{IL}Z_{JK} - G_{JK}Z_{IL} + G_{JL} Z_{IK} \right) \nonumber \\
%& + C_{IJKL} \,,
%\end{align}
\begin{align}
\label{Ricci-decomposition}
R_{IJKL}=&\frac{2 R_\mathrm{fs}}{N_\mathrm{field}(N_\mathrm{field}-1)} G_{I[J}G_{L]K} +\frac{2}{N_\mathrm{field}-2}\left(G_{I[K} Z_{L]J} - G_{J[K}Z_{L]I} \right) + C_{IJKL} \,,
\end{align}
where $R_\mathrm{fs}$ is the Ricci scalar curvature of the field space, 
obtained as the trace of the Ricci tensor $R_{IJ}$, 
$Z_{IJ}=R_{IJ}-\frac{1}{N_\mathrm{field}}R_\mathrm{fs}G_{IJ}$ is the traceless Ricci tensor and $C_{IJKL}$ is the Weyl tensor of the field space.
The brackets denote anti-symmetrization over two indices, for example $G_{I[J}G_{L]K}=(G_{IJ}G_{LK}-G_{IL}G_{JK})/2$.
In the two-field case, $N_\mathrm{field}=2$, both the traceless Ricci tensor $Z$ and the Weyl tensor $C$ vanish, hence only the first term in~\eqref{Ricci-decomposition} remains and we recover the standard result that the geometry of two-dimensional manifolds is completely defined by the Ricci scalar only. Contracting indices of the Riemann tensor with the adiabatic and entropic directions, one finds for the mass contribution:
\begin{align}
\label{Ricci-decomposition2}
R_{\alpha\sigma\beta\sigma}=&\underbrace{\overbrace{\underbrace{\frac{R_\mathrm{fs}}{N_\mathrm{field}(N_\mathrm{field}-1)}\delta_{\alpha\beta}}_{N_\mathrm{field}=2}+ \frac{1}{N_\mathrm{field}-2} \left(Z_{\alpha \beta} + \delta_{\alpha \beta} Z_{\sigma \sigma} \right)}^{N_\mathrm{field}=3} + C_{\alpha \sigma \beta \sigma}}_{N_\mathrm{field} \geqslant  4} \,.
\end{align}
Let us study the effects of the dimension of the field space. When $N_\mathrm{field}=2$, we recover the usual geometrical contribution to the entropic mass $m_s^2 \ni \epsilon R_\mathrm{fs} H^2 \Mp^2$~\cite{GrootNibbelink:2001qt, Langlois:2008mn, Turzynski:2014tza}.
For $N_\mathrm{field}=3$, the Weyl tensor still vanishes but the traceless Ricci tensor is non-zero \textit{a priori} and $m^2_{\alpha\beta} \ni 2 \epsilon H^2 \Mp^2 \left[ \delta_{\alpha\beta} \left( R_\mathrm{fs}/6 + Z_{\sigma \sigma} \right) + Z_{\alpha\beta} \right] $.
The non-diagonal piece of this quantity constitutes a non-trivial geometry-dependent mixing term between different entropic perturbations.
Lastly, for $N_\mathrm{field} \geqslant 4$ one should take into account the Weyl contribution to the Riemann tensor which contains all the geometrical information that is not in the Ricci tensor.

\paragraph*{Linear equations of motion.} The linear equations of motion for the system read:
\begin{align}
\label{linear-eom-zeta}
\mathcal{E}_\zeta=&-\frac{1}{2 a^3 \epsilon \Mp^2 }\frac{\delta S^{(2)}}{\delta \zeta}=0  \\
=&\frac{1}{a^3 \epsilon \Mp^2}\frac{\rm d}{\mathrm{d}t}\left(a^3\epsilon \Mp^2 \dot{\zeta} +a^3 \sqrt{2 \epsilon} \Mp\omega_1 \mathcal{F}^1\right)-  \frac{\partial^2\zeta}{a^2}= \frac{\partial^2}{a^2 \epsilon} \left(\dot{\chi}+H\chi-\epsilon\zeta\right)  \nonumber \,, \\
\label{linear-eom-F}
\mathcal{E}_{\alpha}=&-\frac{1}{a^3} \frac{\delta S^{(2)}}{\delta \mathcal{F}^\alpha}=0 \\
=&\ddot{\mathcal{F}}_\alpha+\left( 3H \delta_ {\alpha \beta} + 2 \Omega_{\alpha\beta}  \right)\dot{\mathcal{F}}^\beta+\left(m_{\alpha\beta}^2 - \delta_{\alpha\beta} \frac{\partial^2}{a^2}+ 3H \Omega_{\alpha \beta} + \dot{\Omega}_{\alpha \beta} \right) \mathcal{F}^\beta \nonumber \\
& - 2 \delta_{\alpha1}  \sqrt{2 \epsilon} \Mp\omega_1\dot{\zeta} \nonumber \,.
\end{align}
These quantities vanish for on-shell linear perturbations, as well as for perturbations in the interaction picture which are evolved with the quadratic action exactly, even when cubic interactions are perturbatively included. 
%Interestingly, although slightly different notations are used, our general second order Lagrangian~\eqref{L2} coincides with the one found in the more particular context of holographic multifield inflation~\cite{Ach_carro_2019}, where the motivation was to study the eigenvalues of the entropic mass matrix.
Our linear equations of motion~\eqref{linear-eom-zeta}-\eqref{linear-eom-F} also match the ones that were already derived in the special case of $N_\mathrm{field}-1=2$ entropic fields in~\cite{C_spedes_2013}.
Strikingly, only the first entropic perturbation couples linearly to the adiabatic curvature perturbation through the mixing parameter $\omega_1$, while the entropic sector mixes when $N_\mathrm{field}\geqslant 3$ both due to the rotation of the entropic frame described by the matrix $\Omega$, and the non-diagonal pieces of the entropic mass matrix (which themselves can come either from the potential, the rotation of the entropic frame or the geometry of the field space).
It is worth noting that attempts to diagonalise the entropic mass matrix and use its eigenvectors to define another entropic frame, must be considered with care as this would necessarily result in more mixing terms between this new entropic sector and $\zeta$.

\paragraph*{Linear stability of the background trajectory.} In the two-field context, it is known that even with a positive Hessian of the potential, the background trajectory can be unstable due to the negative geometrical contribution to the mass of entropic perturbations on large scales, for hyperbolic-like field spaces~\cite{Renaux-Petel:2015mga}.
Here we explain which effects could similarly destabilise a seemingly stable background trajectory (\textit{i.e.} with a covariant Hessian of the potential $V_{;\alpha\beta}$ with only positive eigenvalues) in a more general $N_\mathrm{field}$ context. For this purpose, it is necessary to look at the super-Hubble behaviour of linear perturbations.
Actually, the adiabatic curvature perturbation is fed only by the first entropic perturbation as one finds $\dot{\zeta} = - \sqrt{\frac{2}{\epsilon}} \frac{\omega_1}{\Mp} \F^1+  \mathcal{O}\left(\frac{k^2}{a^2}\right)$, which results in an effective mass matrix for the entropic sector on large scales: $m_{\mathrm{eff},\alpha\beta}^2=m_{\alpha\beta}^2+4 \delta_{\alpha1}\delta_{\beta1} \omega_1^2$.
Thus, as in the $N_\mathrm{field}=2$ case, the bending of the trajectory parameterized by $\omega_1$ eventually has a positive contribution to the effective mass matrix in the first entropic direction, and stabilises the trajectory. However the rotation of the entropic orthonormal basis, parameterized by $\Omega$, can have a destabilising effect when $N_\mathrm{field} \geqslant 3$.
First, the effective mass matrix has a contribution $ \Omega_\alpha{}^\gamma \Omega_{\gamma \beta}$ which, as we saw from~\eqref{Omega2} only has negative eigenvalues, and that could overcome the other terms in $m^2_{\mathrm{eff},\alpha\beta}$\footnote{Note that it was argued that in the limit dubbed ``extreme turning", roughly consisting of a very large turning rate $\omega_1$ (see Ref.~\cite{Aragam:2020uqi} for a more precise statement), the accelerations perpendicular to both the adiabatic and the first entropic directions should be negligible, thus bounding the possible values for the matrix elements $\Omega_{\alpha\beta}$.}.
Secondly, as can be seen from the large-scale EoM for the entropic perturbations,
\begin{align}
\label{eom-large-scales}
\ddot{\mathcal{F}}_\alpha+\left( 3H \delta_ {\alpha \beta} + 2 \Omega_{\alpha\beta}  \right)\dot{\mathcal{F}}^\beta+\left(m_{\mathrm{eff},\alpha\beta}^2 + 3H \Omega_{\alpha \beta} + \dot{\Omega}_{\alpha \beta} \right) \mathcal{F}^\beta = \mathcal{O}\left(\frac{k^2}{a^2} \right) \,,
\end{align}
the effective mass matrix is not the only relevant parameter that determines the stability of the background trajectory: the mixing of entropic perturbations via the rotation matrix $\Omega$ (defined in Eq.~\eqref{cov-rate-matrix})  could result in complicated dynamics worth investigating\footnote{In this spirit, note that Refs.~\cite{Christodoulidis:2019mkj, Christodoulidis:2019jsx} casted a doubt on the only use of the sign of the effective mass matrix of entropic perturbations on large scales in time-dependent setups, to conclude about the stability of a given background trajectory even when $N_\mathrm{field}=2$. It would be interesting to extend this analysis to the present case of any number of fields.}.
Last but not least, it would be interesting to study under which conditions the geometrical term $\propto R_{\alpha\sigma\beta\sigma}$ in $m_{\mathrm{eff},\alpha \beta}^2$ is negative beyond the contribution from the Ricci scalar: how do the traceless Ricci tensor and the Weyl tensor affect the sign of the effective mass matrix of entropic perturbations when $N_\mathrm{field}\geqslant 3$? We leave this interesting questions for future works.

\paragraph*{Hamiltonian formalism.} For completeness, we also show the expressions for the linear canonical momenta of our $N_\mathrm{field}$ variables,
\begin{align}
p_\zeta&= \frac{\delta S }{\delta \dot{\zeta}}=2a^3 \Mp^2 \left( \epsilon\dot{\zeta}+ \sqrt{2\epsilon}  \frac{\omega_1}{\Mp}\,\F^1\right)  = 2a \Mp^2 \partial^2 \chi  \,, \\
p_{\alpha}&= \frac{\delta S}{\delta  \dot{\F}^\alpha}= a^3 \left( \dot{\F}_\alpha + \Omega_{\alpha \beta}  \F^\beta \right) \,,
\end{align}
as well as the corresponding second-order Hamiltonian: 
\begin{align} \label{H2}
\mathcal{H}^{(2)}= &a^3 \left\{\frac{1}{4\epsilon \Mp^2}\tilde{p}_\zeta^2 + \epsilon \Mp^2 \frac{(\partial \zeta)^2}{a^2} - \sqrt{\frac{2}{\epsilon}} \frac{\omega_1}{\Mp} \delta_{\alpha1} \mathcal{F}^\alpha \tilde{p}_\zeta     \right.  \ \\
& \left.  + \frac{1}{2} \left[ \tilde{p}_{\alpha}^2 - 2 \Omega^\alpha{}_{\beta}\tilde{p}_{\alpha}\F^\beta + \frac{(\partial \mathcal{F}^\alpha)^2}{a^2} + \left( m_{\alpha\beta}^2 - \Omega_{\alpha}{}^{\gamma} \Omega
_{\gamma\beta} + 4\delta_{\alpha1}\delta_{\beta 1} \omega_1^2 \right) \mathcal{F}^\alpha \mathcal{F}^\beta  \right] \right\}\,, \nonumber
\end{align}
where we rescaled momenta, $p_\zeta \rightarrow \tilde{p}_\zeta=p_\zeta/a^3$ and   $p_{\alpha} \rightarrow \tilde{p}_{\alpha}=p_{\alpha}/a^3$. Note that an interesting choice of free Hamiltonian to define interaction picture fields, could be this second-order Hamiltonian  but without the mixing terms $\propto \omega_1$ and $\propto \Omega_{\alpha \beta}$.
These mixing terms could then be treated as (quadratic) interactions in a perturbative treatment and in the spirit of Quasi-Single-Field (QSF) inflation~\cite{Chen:2009we,Chen:2009zp,Noumi:2012vr}, but we leave this interesting application for future work, and rather consider the full quadratic Hamiltonian as the free one, so that the interaction picture fields verify the EoM~\eqref{linear-eom-zeta}-\eqref{linear-eom-F} that include mixings non-perturbatively.

\section{Genuine sizes of cubic interactions}
\label{sec3}

We now turn to the computation of the third-order action in the comoving gauge, for any number of scalar fields with kinetic and potential interactions, including the backreaction of the metric and without assuming special features of the background dynamics such as slow-roll.
This corresponds essentially to the $N_\mathrm{field}$-generalisation of the famous Maldacena's calculation of the bispectrum in canonical single-field inflation~\cite{Maldacena:2002vr} and of our previous work in the two-field case~\cite{Garcia-Saenz:2019njm}.
Indeed, as we explained at length in this article, the naive cubic action obtained from expanding the total action, does not render manifest the sizes of interactions and should be manipulated with uses of integrations by parts and linear equations of motion to remove second-order time derivatives.
These manipulations are allowed even in the cubic action, since the fields' perturbations are in the interaction picture and evolve according to the full quadratic Hamiltonian only.
We will thus discard all contributions proportional to the linear EoM, $\mathcal{E}_\zeta$ and $\mathcal{E}_{\alpha}$.
However, due to the special feature of the ``in-in" formalism that consists in computing cosmological correlators at a given time-boundary of interest, and contrary to ``in-out" amplitudes in particle physics, total time derivatives in the cubic action play a role in the general computation of three-point functions and should be kept for consistency throughout the calculation. The reader that is not interested in the derivation can directly jump to the final result~\eqref{L3-final}-\eqref{boundary-final}.

\subsection{Brute-force expansion}

First, one expands all contributions to the full action~\eqref{action-ADM} (supplemented by the GHY boundary term) to cubic order in perturbations, and obtains after uses of the background EoM~\eqref{back-eom-localframe} and of the Friedmann constraints~\eqref{back-Friedmann}:
\begin{equation}
\mathcal{L}^{(3)}=\mathcal{L}_{\rm ini}^{(3)}(\zeta,\theta)+\mathcal{L}_{\rm ini}^{(3)}(\zeta,\F^\alpha)+\mathcal{D}_0 \,.
\end{equation}
The first part, $\mathcal{L}_{\rm ini}^{(3)}(\zeta,\theta)$, comes (mostly) from the gravity sector described by the EH action in the comoving gauge and depends on the entropic sector only through the constraint equation relating $\theta$ to $\F^1$. It reads
\begin{align}
\label{L3simple-noF}
\mathcal{L}_{\rm ini}^{(3)}(\zeta,\theta)=a^3\Mp^2 & \left[ \epsilon \left(3\zeta-\frac{\dot{\zeta}}{H}\right)\dot{\zeta
}^2   -\frac{\epsilon}{a^2} \zeta \left(\partial \zeta\right)^2
+ \frac{1}{2a^4}\left(3\zeta-\frac{\dot{\zeta}}{H} \right) \left(\partial_{ij}\theta\partial_{ij}\theta - (\partial^2\theta)^2\right) \right. \nonumber \\
& \left. -\frac{2}{a^4} \partial^2 \theta \left(\partial\theta\right) 
\left(\partial \zeta\right)  \right] \,,
\end{align}
exactly like in the two-field case~\cite{Garcia-Saenz:2019njm}. The second part, $\mathcal{L}_{\rm ini}^{(3)}(\zeta,\F^\alpha)$, originates from the matter action in the comoving gauge and has no single-field counterpart by definition: 
\begin{align}
\label{L3simple-withF}
\mathcal{L}_{\rm ini}^{(3)}\left(\zeta,\mathcal{F}^\alpha\right)=&\frac{a^3}{2}\left[\left(3\zeta-\frac{\dot{\zeta}}{H}\right)\left(\dot{\mathcal{F}}^\alpha\right)^2+2 \sqrt{2\epsilon}\omega_1 \Mp \mathcal{F}^1 \dot{\zeta} \left(6\zeta- \frac{\dot{\zeta}}{H} \right)-3m_{\alpha\beta}^2 \mathcal{F}^\alpha \mathcal{F}^\beta\zeta \right. \nonumber \\
& \left. +2 \Omega_{\alpha\beta} \dot{\mathcal{F}}^\alpha \mathcal{F}^\beta \left(3\zeta- \frac{\dot{\zeta}}{H} \right) - \mathcal{F}^\alpha \mathcal{F}^\beta \frac{\dot{\zeta}}{H}\left(m_{\alpha\beta}^2 - 2 \Omega_\alpha{}^\gamma \Omega_{\gamma\beta}+2 \delta_{\alpha 1} \delta_{\beta 1} \omega_1^2 \right. \right. \nonumber \\ 
& \left. \left. - 4 \epsilon H^2 \Mp^2 R_{ \alpha \sigma \beta \sigma} \right) 
 + \frac{4}{3}\sqrt{2\epsilon} H \Mp R_{\alpha\beta\gamma\sigma} \dot{\mathcal{F}}^\alpha\mathcal{F}^\beta \mathcal{F}^\gamma \right. \nonumber \\
 &  \left.  -\frac{\mathcal{F}^\alpha \mathcal{F}^\beta \mathcal{F}^\gamma}{3} \left( V_{;\alpha\beta\gamma}   - 4 \sqrt{2 \epsilon} H \Mp \left(\omega_1 \delta_{\alpha 1} R_{ \beta \sigma \gamma \sigma} + \Omega^\delta{}_\alpha R_{\delta\beta\gamma\sigma} \right) + 2 \epsilon H^2 \Mp^2 R_{\alpha \sigma \beta \sigma; \gamma} \right) \right. \nonumber \\
& \left. -2\dot{\mathcal{F}}^\alpha\frac{\partial_i\mathcal{F}^\alpha \partial_i \theta}{a^2} - \frac{(\partial\mathcal{F}^\alpha)^2}{a^2} \left(\zeta+\frac{\dot{\zeta}}{H} \right)- 2 \frac{\partial_i\mathcal{F}^\alpha \partial_i \theta}{a^2} \Omega_{\alpha \beta} \mathcal{F}^\beta \right] \,.
\end{align}
Note that it obviously contains additional multifield interactions compared to our previous work~\cite{Garcia-Saenz:2019njm}, which are proportional to the purely entropic anti-symmetric matrix $\Omega$ or to three contractions of the field-space Riemann tensor along entropic directions.
Indeed, this last quantity is vanishing in the two-field case where there is only one entropic direction, as can be seen from the symmetries of the Riemann tensor, or equivalently by the observation that $R_{\alpha\beta\gamma\sigma}=\frac{1}{N_\mathrm{field}-2}\left(\delta_{\alpha\gamma} Z_{ \beta \sigma}-\delta_{\beta\gamma}Z_{\alpha \sigma} \right) + C_{\alpha\beta\gamma\sigma}$ contains geometrical information beyond $R_\mathrm{fs}$ only.
Eventually, at this stage the time-boundary term simply reads:
\begin{equation}
\mathcal{D}_0=-  \Mp^2 \partial_t \left[9Ha^3\zeta^3 - \frac{a}{H} \zeta (\partial \zeta)^2 \right] \,,
\end{equation}
and comes from a few manipulations to get the form of $\mathcal{L}_{\rm ini}^{(3)}(\zeta,\theta)$ that is displayed in~\eqref{L3simple-noF}.

\subsection{Rendering the sizes of interactions manifest}

Using integrations by parts and the linear EoM~\eqref{linear-eom-zeta}-\eqref{linear-eom-F} to remove second-order time derivatives, but keeping time-boundary terms, it is possible to simplify and render explicit the sizes of cubic interactions.
We split this calculation in two steps, one for each of the cubic order Lagrangians showed in~\eqref{L3simple-noF} and~\eqref{L3simple-withF}.

\paragraph{Lagrangian $\mathcal{L}_{\rm ini}^{(3)}(\zeta,\theta)$.}
This part of the computation is exactly the same as in the $N_\mathrm{field}=2$ case, with the formal replacements $\mathcal{F}\rightarrow \mathcal{F}^1$ and $\eta_\perp \rightarrow \omega_1/H$ in the various steps of~\cite{Garcia-Saenz:2019njm}, coming from the constraint equation for $\theta$ and the linear EoM for $\zeta$.
We refer the interested reader to this article and only quote the final result:
\begin{align}
\label{L3-result-noF}
\mathcal{L}_{{\rm ini}}^{(3)}(\zeta,\theta)&= a^3 \Mp^2\left[ \epsilon (\epsilon-\eta) \dot{\zeta}^2\zeta+ \epsilon (\epsilon+\eta)\zeta \frac{\left(\partial \zeta \right)^2}{a^2}  +
\left(\frac{\epsilon}{2}-2\right)\frac{1}{a^4} \left(\partial\zeta\right)\left(\partial\chi\right) \partial^2 \chi + \frac{\epsilon}{4 a^4} \partial^2 \zeta ( \partial \chi)^2 \right] \nonumber \\
&   +a \sqrt{2 \epsilon} \frac{\omega_1}{H} \Mp \F^1 (\partial\zeta)^2 - \frac{2}{H} \dot{\zeta} \zeta \partial_t \left(a^3 \sqrt{2 \epsilon}  \omega_1 \Mp \F^1 \right)+\mathcal{D}_1,
\end{align}
with
\begin{align}
\mathcal{D}_{1}=&- \Mp^2 \partial_t  \bigg[ a^3 \frac{\epsilon}{H} \zeta \dot{\zeta}^2+ a \frac{\epsilon}{H}\zeta (\partial\zeta)^2 + \frac{1}{6aH^3} \zeta \left( \zeta_{,ij}\zeta_{,ij}-(\partial^2\zeta)^2 \right) - \frac{1}{2aH^2} \zeta\left(\zeta_{,ij}\chi_{,ij}-\partial^2\zeta \partial^2\chi\right) \nonumber \\
&+ \frac{1}{2aH} \zeta \left(\chi_{,ij}\chi_{,ij}- \left(\partial^2\chi \right)^2 \right)   \bigg] \,,
\end{align}
and we recall that $\chi$ is given by~\eqref{def-chi}.

\paragraph{Lagrangian $\mathcal{L}_{\rm ini}^{(3)}\left(\zeta,\mathcal{F}^\alpha\right)$.}

Let us manipulate three building blocks of this purely multifield Lagrangian, which will eventually combine to simplify interactions and render their sizes more explicit. First, combining the first term in the first line with the three terms of the last line of~\eqref{L3simple-withF}, using integrations by parts and the linear EoM for $\zeta$ and $\F^\alpha$, one finds
\begin{align}
\label{equality1}
\frac{a^3}{2}&\left(\dot{\F}^\alpha\right)^2\left(3 \zeta - \frac{\dot{\zeta}}{H} \right) -a \dot{\mathcal{F}}^\alpha \partial_i\mathcal{F}^\alpha \partial_i \theta - \frac{a}{2} \left(\partial\mathcal{F}^\alpha\right)^2 \left(\zeta+\frac{\dot{\zeta}}{H} \right) -a \partial_i\mathcal{F}^\alpha \partial_i \theta \Omega_{\alpha \beta} \mathcal{F}^\beta \nonumber \\
&=  a^3 \left[ \frac{\epsilon}{2} \zeta \left(\dot{\F}^\alpha \right)^2 - \frac{\zeta}{H} \dot{\F}^\alpha \F^\beta \left( m^2_{\alpha \beta} + 3 H \Omega_{\alpha\beta} + \dot{\Omega}_{\alpha \beta} \right) + \frac{2}{H} \sqrt{2\epsilon} \omega_1 \Mp \zeta \dot{\zeta} \dot{\F}^1 \right] \nonumber  \\
& \, \, \, + a \left[ \frac{\epsilon}{2} \zeta \left(\partial \F^\alpha\right)^2 - \left(\partial\F^\alpha \right)\left(\partial\chi\right)\left(\dot{\F}^\alpha+\Omega_{\alpha \beta} \F^\beta \right) + \frac{1}{H} \left(\partial\F^\alpha \right)\left(\partial \zeta \right)\Omega_{\alpha \beta} \F^\beta \right] \nonumber \\
& - \partial_t \left[ \frac{a^3}{2H} \zeta \left( \left(\dot{\F}^\alpha \right)^2 + \frac{\left(\partial \F^\alpha \right)^2 }{a^2} \right)  \right] \,.
\end{align}
It is clear from this result that, for example, the right size of the $\zeta$-$\left(\partial_\mu\F^\alpha \right)^2$ interactions is more suppressed (by a factor of $\epsilon$ in this case) than what it seems when looking at the brute-force initial cubic Lagrangian. Let us proceed and look at a second block, made of the second and third terms of the first line in~\eqref{L3simple-withF}, to which one adds the first term in the second line of the result coming from~\eqref{L3-result-noF} as well as two terms in the first line of the previous equality~\eqref{equality1}:
\begin{align}
& a^3  \sqrt{2\epsilon}\omega_1 \Mp \F^1 \dot{\zeta}\left( 6 \zeta - \frac{\dot{\zeta}}{H} \right) - \frac{3}{2} a^3 m^2_{\alpha\beta} \zeta \F^\alpha \F^\beta - \frac{2}{H} \dot{\zeta} \zeta \partial_t\left(a^3 \sqrt{2 \epsilon} \omega_1 \Mp \F^1 \right)  \nonumber \\
 & + a^3 \frac{2}{H} \sqrt{2 \epsilon} \omega_1 \Mp \zeta \dot{\zeta} \dot{\F}^1 - \frac{a^3}{H}m^2_{\alpha\beta} \zeta \dot{\F}^\alpha \F^\beta \nonumber \\
 &=   a^3 \left[  \frac{1}{2} m^2_{\alpha\beta}  \frac{\dot{\zeta}}{H}\F^\alpha \F^\beta + \frac{1}{2}\left(\epsilon m^2_{\alpha\beta} + \frac{\partial_t\left(m^2_{\alpha\beta}\right)}{H} \right) \zeta \F^\alpha \F^\beta - \sqrt{2 \epsilon} \omega_1 \Mp \left( \frac{\dot{\zeta}^2}{H} \F^1 + 2 \dot{\zeta} \zeta \F^1 \left( \frac{\eta}{2} + u_1 \right) \right)  \right] \nonumber \\
 & - \partial_t \left( m^2_{\alpha \beta } \frac{a^3}{2H} \zeta \F^\alpha \F^\beta \right) \,.
\end{align}
Next, we use
\begin{align}
&-\frac{a^3}{H} \Omega^\gamma{}_\alpha \Omega_{\gamma \beta} \dot{\zeta} \F^\alpha \F^\beta + a^3 \Omega_{\alpha \beta} \dot{\F}^\alpha \F^\beta \left(3 \zeta - \frac{\dot{\zeta}}{H} \right) = a^3 \left[ \left( \left(3+\epsilon\right) \Omega_{\alpha\beta} + \frac{ \dot{\Omega}_{\alpha\beta} }{H} \right) \zeta \dot{\F}^\alpha\F^\beta \right. \nonumber \\
& \left. + \left( \epsilon \Omega^\gamma{}_\alpha  +  \frac{\dot{\Omega}^\gamma{}_\alpha}{H} - \frac{m^2_{\gamma\alpha}}{H}\right) \Omega_{\gamma \beta}\zeta \F^\alpha \F^\beta + \frac{2}{H} \sqrt{2\epsilon} \omega_1 \Mp  \Omega_{1\alpha} \dot{\zeta} \zeta  \F^\alpha \right]  + \frac{a}{H}\Omega_{\alpha\beta} \zeta  \partial^2 \F^\alpha \F^\beta \nonumber \\
& - \partial_t \left( \frac{a^3}{H} \Omega_{\alpha\beta} \zeta \dot{\F}^\alpha \F^\beta + \frac{a^3}{H}  \Omega^\gamma{}_\alpha \Omega_{\gamma \beta} \zeta \F^\alpha \F^\beta \right) \,.
\end{align}
Collecting these three blocks, being careful not to double count some terms, and adding them to the rest of the Lagrangian that has not been manipulated, one finds many cancellations and arrives at the final result:
\begin{align}
\label{L3-final}
\mathcal{L}^{(3)}&=M_p^2a^3\left[\epsilon(\epsilon-\eta)\dot{\zeta}^2\zeta+\epsilon (\epsilon+\eta) \zeta \frac{(\partial \zeta)^2}{a^2}+\left(\frac{\epsilon}{2}-2\right)\frac{1}{a^4} (\partial \zeta
) (\partial \chi ) \partial^2 \chi + \frac{\epsilon}{4a^4} \partial^2 \zeta (\partial \chi )^2 \right] \nonumber \\
&+a^3\left\{ \sqrt{2\epsilon} \omega_1 \Mp \left[  \frac{\F^1}{H} \left( \frac{\left( \partial \zeta \right)^2}{a^2} - \dot{\zeta}^2- \dot{\zeta} \zeta H \left( \eta +2u_1 \right) \right) + 2 \frac{ \Omega_{1\alpha} }{H} \dot{\zeta} \zeta \F^\alpha \right]  \right. \nonumber \\
& \left. + \left[ \frac{\epsilon}{2} m_{\alpha\beta}^2 +  \frac{\dot{\left(m_{\alpha\beta}^2\right)}}{2H} + \Omega_{\gamma \beta} \left( \epsilon \Omega^\gamma{}_\alpha  +  \frac{\dot{\Omega}^\gamma{}_\alpha}{H} - \frac{m^2_{\gamma\alpha}}{H}\right)  \right] \zeta \mathcal{F}^\alpha \mathcal{F}^\beta + \epsilon \Omega_{\alpha\beta} \zeta \dot{\F}^\alpha \F^\beta \right.  \nonumber  \\
 &  \left. + \left(   2 \epsilon H^2 \Mp^2 R_{\alpha \sigma \beta \sigma} - \omega_1^2\delta_{\alpha 1} \delta_{\beta 1 }  \right)\frac{\dot{\zeta}}{H}\F^\alpha\F^\beta + \frac{1}{2}\epsilon \zeta \left(\left( \dot{\mathcal{F}}^\alpha\right)^2+ \frac{\left(\partial \mathcal{F}^\alpha \right)^2}{a^2}\right) \right.  \\
 & \left.   - \frac{1}{a^2}  \left(\partial \mathcal{F}^\alpha \right) (\partial \chi) \left( \dot{\mathcal{F}}^\alpha + \Omega_{\alpha \beta} \F^\beta \right) + \frac{2}{3} \sqrt{2 \epsilon} H \Mp R_{\alpha \beta \gamma \sigma} \dot{\F}^\alpha \F^\beta \F^\gamma \right. \nonumber \\
 & \left . -\frac{1}{6} \left( V_{;\alpha\beta\gamma}   - 4 \sqrt{2 \epsilon} H \Mp \left(\omega_1 \delta_{\alpha 1} R_{ \beta \sigma \gamma \sigma} + \Omega^\delta{}_\alpha R_{\delta\beta\gamma\sigma} \right) + 2 \epsilon H^2 \Mp^2 R_{\alpha \sigma \beta \sigma; \gamma} \right)  \mathcal{F}^\alpha \mathcal{F}^\beta \mathcal{F}^\gamma \right\} \nonumber \\
& + \mathcal{D} \nonumber \,,
\end{align}
where we defined the logarithmic derivative of the first bending as $u_1= \dot{\omega}_1/\left(H \omega_1 \right)$, and with the final boundary term
\begin{align}
\label{boundary-final}
\mathcal{D}=-\partial_t&\left\{  \Mp^2 \left[  9Ha^3\zeta^3 - \frac{a}{H} \zeta (\partial \zeta)^2  + a^3 \frac{\epsilon}{H} \zeta \dot{\zeta}^2+ a \frac{\epsilon}{H}\zeta (\partial\zeta)^2 + \frac{1}{6aH^3} \zeta \left( \zeta_{,ij}\zeta_{,ij}-(\partial^2\zeta)^2 \right) \right. \right. \nonumber \\
& \left. \left. - \frac{1}{2aH^2} \zeta\left(\zeta_{,ij}\chi_{,ij}-\partial^2\zeta \partial^2\chi\right) + \frac{1}{2aH} \zeta \left(\chi_{,ij}\chi_{,ij}- \left(\partial^2\chi \right)^2 \right)  \right]  \right.   \\
& \left. + \frac{a^3}{2H} \zeta \left[ \left(\dot{\F}^\alpha \right)^2 + \frac{\left(\partial \F^\alpha \right)^2 }{a^2} \right] + m^2_{\alpha \beta } \frac{a^3}{2H} \zeta \F^\alpha \F^\beta + \frac{a^3}{H} \Omega_{\alpha\beta} \zeta \dot{\F}^\alpha \F^\beta + \frac{a^3}{H}  \Omega^\gamma{}_\alpha \Omega_{\gamma \beta} \zeta \F^\alpha \F^\beta  \right\} \nonumber \,,
\end{align}
whose local (in time) contribution to the three-point functions can be easily computed.

\section{Single-field effective theory of fluctuations}
\label{sec4}

Even when the background inflationary trajectory is genuinely multifield, it may be possible to describe perturbations as a single-field system only.
This so-called effective single-field theory of fluctuations is justified when entropic perturbations $\F^\alpha$ are heavy enough to be adiabatically expressed in terms of the adiabatic perturbation $\zeta$ to which they are mixing, and thus integrated out of the multifield theory.
Within its regime of validity, this tool is extremely powerful to compute efficiently the statistics of the observable curvature perturbation as in single-field inflation.
The multifield dynamics is however encoded in the parameters setting the free propagation (through $c_s^2$ the speed of sound) and the interactions of $\zeta$ (the to-be-found parameter $A$ for cubic interactions, and others for higher-order ones), thus obviously affecting the theoretical predictions for cosmological observables~\cite{Tolley:2009fg,Achucarro:2012sm, Gwyn:2012mw, Burgess:2012dz, Cespedes:2013rda, Garcia-Saenz:2019njm}.
In this section, we remind that the first bending $\omega_1$ plays a special role: if vanishing, the entropic sector entirely decouples from the adiabatic perturbation and the resulting single-field effective theory is unaffected by multifield effects, both at the level of linear fluctuations and Non-Gaussianities.
However, as soon as $\omega_1 \neq 0$, the speed of sound is reduced, $c_s^2 < 1$, and cubic interactions are enhanced compared to the purely single-field case.
As expected, we show that the resulting single-field cubic action can always be recast in the form of $P(X,\phi)$ cubic interactions. In the slow-varying limit, we recover the two leading-order cubic operators known from the decoupling limit of the Effective Field Theory of Inflation (EFToI)~\cite{Creminelli:2006xe,Cheung:2007st}, with a prediction for the Wilson coefficient $A$ coming from the multifield UV completion.
Importantly, $A$ contains contributions from the non-trivial speed of sound $c_s^2$, from third derivatives of the scalar potential, but also from the geometry of the field-space even beyond the scalar curvature: the traceless Ricci tensor and the Weyl tensor matter too.

\subsection{Generalities}

\paragraph*{Integrating out $N_\mathrm{field}-1$ heavy entropic perturbations.}Let us recall that the EoM~\eqref{linear-eom-F} for entropic perturbations $\F^\alpha$ can be written:
\begin{align}
\left(m^2_{\alpha\beta}+ \mathcal{O}^\Omega_{\alpha\beta} -   \Box_{\alpha \beta}  \right)\F^\beta&=2\sqrt{2\epsilon}\delta_{\alpha1} \Mp\omega_1 \dot{\zeta} \,, \\
 \text { with } \Box_{\alpha\beta}&=\delta_{\alpha \beta} \left [ -\frac{\partial^2 }{\partial t^2} - 3H \frac{\partial}{\partial t} + \frac{\partial^2}{a^2} \right] \text{ the d'Alembertian operator} \nonumber \\
\text{ and }  \mathcal{O}^\Omega_{\alpha\beta} &= 2\Omega_{\alpha\beta} \frac{\partial}{\partial t}+\left(3H\Omega_{\alpha\beta} +\dot{\Omega}_{\alpha\beta} \right) \text{ the entropic-mixing operator}  \nonumber \,.
\end{align}
In the limit $m^2_{\alpha \beta}\mathcal{F}^\beta \gg \left( \mathcal{O}^\Omega_{\alpha\beta}  - \Box_{\alpha\beta} \right) \mathcal{F}^\beta $ that we shall call ``heavy entropic perturbations" and whose regime of validity will soon be specified precisely, the equation of motion for $\mathcal{F}^\alpha$ can be solved at leading order (LO) by
\begin{equation}
\label{F-LO}
\mathcal{F}^\alpha_{\mathrm{LO}}=(m^{-2})^\alpha{}_1 2 \sqrt{2 \epsilon}  \Mp \omega_1 \dot{\zeta} \,,
\end{equation}
where $m^{-2}$ is the inverse matrix of $m^2$. Interestingly, only the first column of this inverse matrix is needed, and one can use a famous linear algebra formula applied to the symmetric matrix $m^2$, to rewrite:
\begin{equation}
(m^{-2})^\alpha{}_1 = \frac{\left(\mathrm{Adj}\left(m^2\right)\right)^\alpha{}_1}{\mathrm{Det}\left(m^2\right)} \,,
\end{equation}
where $\mathrm{Adj}\left(m^2\right)={}^t\mathrm{Com}(m^2)$ is the adjugate of the entropic mass matrix, or equivalently its cofactor matrix $\mathrm{Com}(m^2)$ since it is symmetric. Now, plugging into the $N_\mathrm{field}$ action the expression for $\F^\alpha_{\mathrm{LO}}$ in~\eqref{F-LO}, one can find the effective single-field theory for $\zeta$ when heavy entropic perturbations have been integrated out.

\paragraph*{Linear dynamics.} Consistently neglecting kinetic terms of $\F^\alpha$ and its mixings to other entropic modes through $\Omega$, one finds the quadratic effective Lagrangian in terms of $\zeta$ only,
\begin{equation}
\label{EFT-L2}
\mathcal{L}^{(2)}_{\mathrm{EFT}}(\zeta)=a^3 \frac{\epsilon \Mp^2}{c_s^2} \left(\dot{\zeta}^2 - \frac{(\partial \zeta)^2}{a^2} \right) \text{   with   } \frac{1}{c_s^2}-1=4 \omega_1^2 \left(m^{-2}\right)_{11} \,,
\end{equation}
which results in the effective EoM for $\zeta$:
\begin{equation}
\mathcal{E}_{\zeta,\mathrm{EFT}}^\mathrm{LO}=\frac{1}{a^3 \epsilon}\frac{\mathrm{d}}{\mathrm{d} t} \left(a^3  \frac{\epsilon}{c_s^2}\dot{\zeta}\right) - \frac{\partial^2 \zeta}{a^2}=0 \,,
\end{equation}
showing that a deviation from a speed of sound $c_s^2=1$ is possible only if the first bending parameter $\omega_1$ is non-zero. Indeed, $\omega_1$ is the single parameter that controls the mixing of the whole entropic sector (which has its own mixings and interactions) with the adiabatic curvature perturbation.
%As we shall see now by computing the effective cubic interactions, this statement holds at next order in the bispectrum and actually at any order.
Note however that $\left(m^{-2}\right)_{11}$ can be approximated by $1/\left(m^{2}\right)_{11}$ only if the matrix $m^2$ is mostly diagonal, and that otherwise the whole $N_\mathrm{field}$ system contributes to the effectively single-field linear dynamics through $\left(m^{-2}\right)_{11}$ whenever $\omega_1 \neq 0$. Before turning to the computation of the effective single-field interactions in the cubic action, let us make more precise the conditions under which the procedure of integrating out entropic fluctuations is justified.

\paragraph{Regime of validity.} Assuming that $\F^\alpha$ can be simply expressed as $\F_\mathrm{LO}^\alpha$ displayed in Eq.~\eqref{F-LO} amounts to neglecting higher-order corrections. Indeed, the full  ``non-perturbative" result for $\F^\alpha$ is rather
\begin{equation}
\F^\alpha=  \left\{ \left( m^{-2} \right)  \sum_{i=0} \left[ m^{-2} \left( \Box - \mathcal{O}^\Omega \right) \right]^i \right\}_{\alpha 1}  2 \sqrt{2\epsilon}\Mp \omega_1 \dot{\zeta}  \,,
\end{equation}
where higher powers of $i$ are expected to be more suppressed in the regime of validity of the single-field EFT. As a consistency check, we must require that the next-to-leading-order (NLO) term $\F_\mathrm{NLO}^\alpha$ ($i=1$) is smaller than the leading-order one $\F_\mathrm{LO}^\alpha$ ($i=0$). Computing the NLO term $\F^\alpha_\mathrm{NLO}= \left\{ \left(m^{-2} \right) \left[ m^{-2} \left( \Box - \mathcal{O}^\Omega\right)\right] \right\}_{\alpha 1}  2 \sqrt{2\epsilon}\Mp \omega_1 \dot{\zeta} $ requires in particular to compute 
\begin{align}
\Box_{\alpha\beta}\left(2 \sqrt{2\epsilon}\Mp \omega_1 \dot{\zeta}  \right)= & \delta_{\alpha\beta} 2 \sqrt{2\epsilon}\Mp \omega_1 \left\{ \left(1-c_s^2\right)\frac{\partial^2\dot{\zeta}}{a^2} +c_s^2\left(2-2u_1-4s\right) H \frac{\partial^2\zeta}{a^2} \right.  \\
& \left. +H^2  \dot{\zeta}\left[\left(-3-\frac{\eta}{2}+u_1+2s\right)\left(\epsilon+\frac{\eta}{2}-u_1-2s\right) + \frac{\dot \eta}{2 H} - \frac{\dot{u}_1}{H}-2\frac{\dot s}{H}\right] \right\}    \nonumber \,, \\
\mathcal{O}^\Omega_{\alpha\beta}  \left(2 \sqrt{2\epsilon}\Mp \omega_1 \dot{\zeta}  \right) = &  2 \sqrt{2\epsilon}\Mp \omega_1 \left\{ 2 \Omega_{\alpha\beta} c_s^2 \frac{\partial^2\zeta}{a^2}  + H\dot{\zeta}\left[ \Omega_{\alpha\beta} \left(2u_1 +2s \right) + \frac{\dot{\Omega}_{\alpha\beta}}{H} \right] \right\} \nonumber \,,
\end{align}
where $s=\dot{(c_s^2)}/(2H c_s^2)=(c_s^2-1)\left\{ u_1 + \frac{1}{2H} \frac{\mathrm{d}}{\mathrm{d} t } \left[(m^{-2})_{11}\right]/(m^{-2})_{11}\right\}$ is the logarithmic time derivative of the speed of sound $c_s$, and from which we infer more precise conditions on the multifield parameters for the single-field theory to be valid:
\begin{align}
\label{validity}
\left.
\begin{aligned}
\text{ for } X \in \left(H,\epsilon,\omega_1,c_s \right) , ~~ \mathcal{O}\left[ \left(\frac{\dot{X}}{X}\right)^2 , \frac{\ddot{X}}{X}  \right]\left( m^{-4}\right)^\alpha{}_1 &   \\
\frac{k^2 \left(1-c_s^2\right)}{a^2} \left(m^{-4}\right)^\alpha{}_1 &  \nonumber \\
\left( m^{-4} \dot{\Omega}  \right)^\alpha{}_1 &  \nonumber \\
\left(u_1+s\right)H\left( m^{-4}  \Omega \right)^\alpha{}_1 &  \nonumber \\
\end{aligned}
\right\} \ll \left(m^{-2} \right)^\alpha{}_1 \addtocounter{equation}{1}\tag{\theequation} \,,
\end{align}
as well as less clear ``operators" inequalities: $c_s^2 \left[H \left( m^{-4}\right)^\alpha{}_1  +\left(m^{-4} \Omega\right)^\alpha{}_1  \right] \partial^2 \zeta / a^2 \ll \left(m^{-2}\right)^\alpha{}_1 \dot{\zeta}$.
The first line of~\eqref{validity} means that the first and second derivatives of background quantities must be slowly evolving in units of the entropic squared mass matrix.
Had we considered higher terms in the $i$-expansion, we would have found that even higher derivatives should be small in those units.
Eventually, this means that the whole functions describing the background, and not just their first derivatives, should be slowly evolving during a time scale set by the mass matrix $m^2$.
These adiabaticity conditions are the same as the ones previously found in the case of $N_\mathrm{field}=2$ scalar fields~\cite{Garcia-Saenz:2019njm} and that were first proposed in~\cite{Cespedes:2012hu}, but generalized to a system of any number of entropic fields.
The second line in~\eqref{validity} is important as it sets the UV cutoff of the single-field EFT in terms of wavenumbers.
It states that only wavenumbers below the scale set by the entropic mass are incorporated in the EFT, higher momenta being excluded from this description as they are affected by new physics~\cite{Cremonini:2010ua,Baumann:2011su}.
This makes sense because for higher values of the momenta, the spatial derivatives of the entropic perturbations are becoming larger than the mass terms and the scalar fields should behave as in Bunch-Davies vacuum.
The third and fourth lines in~\eqref{validity} are new requirements coming from the fact that several entropic perturbations were integrated out, and constrain the maximal amount of rotation (and its derivative) of the local entropic basis.
Although the vectorial structure of this equation is such that only the second bending $\omega_2$ seems to be constrained, again it can be shown that for higher values of $i$ we would get constraints on a larger set of bendings and their multiple derivatives.
Thus, from the third line we understand that it is actually the whole matrix $\Omega$ which is constrained to be slowly evolving in units of the entropic mass matrix, and not just the first or the two first bendings and their first derivatives.
The fourth line is more intricate to understand as it mixes derivatives of $\omega_1$ and $c_s$ with the matrix $\Omega$ but it roughly states that the bendings in $\Omega$ can not be too large, a necessary requirement to neglect $\mathcal{O}^\Omega_{\alpha\beta}$ in the EoM for $\F^\alpha$, at leading order.

In the regime where those conditions are satisfied, time derivatives of entropic perturbations,  $\dot{\F}^\alpha_\mathrm{LO}$, which although generically suppressed at the linear level will be needed in the cubic action, read:
\begin{equation}
\label{F-dot-LO}
\dot{\mathcal{F}}^\alpha_\mathrm{LO}=2 \sqrt{2 \epsilon} \Mp \omega_1 \left\{\left(m^{-2}\right)^\alpha{}_1 \left[ \left(-3 - \frac{\eta}{2} + u_1 + 2s \right) H \dot{\zeta} + c_s^2 \frac{\partial^2 \zeta}{a^2} \right]+\frac{\mathrm{d}}{\mathrm{d} t}\left[\left(m^{-2}\right)^\alpha{}_1\right] \dot{\zeta} \right\} \,.
\end{equation}

\subsection{Effective cubic interactions}

We begin by recalling that, in order to derive the effectively single-field cubic interactions, it is sufficient to plug into the full $N_\mathrm{field}$ action the linear constraints of $\F^\alpha$ and its derivative as a function of $\zeta$, \textit{i.e.} quadratic corrections are irrelevant.
Indeed, let us write formally $\F^\alpha\left[\zeta\right]=\F^\alpha_{(1)}\left[\zeta\right]+\F^\alpha_{(2)}\left[\zeta\right]$ where $\F^\alpha_{(1)}$ is linear in $\zeta$ and comes from the second action only, and $\F^\alpha_{(2)}$ is quadratic and comes from the correction due to the cubic action.
Then, up to cubic order in perturbations, one gets
\begin{align}
S_\mathrm{EFT}\left[\zeta\right]&=S^{(2)}\left[\zeta,\F^\alpha_{(1)}\left[\zeta\right]+\F^\alpha_{(2)}\left[\zeta\right] \right] +  S^{(3)}\left[\zeta,\F^\alpha_{(1)}\left[\zeta\right] \right] \\
&= S^{(2)}\left[\zeta,\F^\alpha_{(1)}\left[\zeta\right]\right] + \underbrace{\biggl. \frac{\delta S^{(2)}}{\delta \F^\alpha} \biggr|_{\zeta,\F^\alpha_{(1)}\left[\zeta\right]}}_{=0} \times \F^\alpha_{(2)}\left[\zeta\right] +  S^{(3)}\left[\zeta,\F^\alpha_{(1)}\left[\zeta\right] \right] \nonumber \,,
\end{align}
where the second term in the second line vanishes by virtue of the linear EoM, in a way that only $\F^\alpha_{(1)}$ is relevant. Of course this is true only up to cubic order, and for higher-order interactions, one would need to take into account non-linear corrections to $\F^\alpha_{(1)}$.

\paragraph{General result.}
Performing the integrating-out procedure at the level of the full multifield cubic action~\eqref{L3-final} (but disregarding the single-field boundary terms that give a vanishing contribution to the adiabatic power spectrum on super-Hubble scales), consistently neglecting terms that are not encapsulated by the leading-order expressions~\eqref{F-LO} and~\eqref{F-dot-LO}, yields the following result:
\begin{equation}
\label{L3-EFT}
\mathcal{L}^{(3)}_{{\rm EFT, bulk}}= \Mp^2 \,a^3 \frac{\epsilon}{c_s^2}\left[f_0 c_s^2 \frac{\dot{\zeta}}{H} \frac{\left(\partial\zeta\right)^2}{a^2}+\frac{f_1}{H} \dot{\zeta}^3 + f_2 \dot{\zeta}^2\zeta + f_3 c_s^2 \zeta \frac{\left(\partial\zeta\right)^2}{a^2}  + f_4 \dot{\zeta} \partial_i\partial^{-2}\dot{\zeta}\partial_i\zeta + f_5 \partial^2\zeta (\partial_i \partial^{-2} \dot{\zeta})^2\right] \,,
\end{equation}
with
\begin{align}
f_0&=\left(\frac{1}{c_s^2}-1\right)  \qquad f_1=\left(\frac{1}{c_s^2}-1\right) A \qquad f_2=\epsilon-\eta+2s \nonumber \\
   f_3&=\epsilon+\eta  \qquad \qquad \,  f_4=\frac{\epsilon}{2 c_s^2}(\epsilon-4)  \qquad  \,\,\,\,\,f_5=\frac{\epsilon^2}{4c_s^2}  \,.
\label{fi}
\end{align}
The parameter $A$ encoding multifield effects reads
\begin{align}
\label{A}
A&=-\frac{1}{2}(1+c_s ^2) + \frac{4}{3}(1+2c_s^2) \epsilon H^2 M_p^2 \left(m^{-2}\right)_{11} R_{ m \sigma m \sigma}   \\ 
&- \frac{\kappa}{6}(1-c_s^2) \left(m^{-2}\right)_{11} \biggl[ V_{; m m m } + 2 \epsilon H^2 M_p^2 R_{m \sigma m \sigma ; m} \bigr. \nonumber \\
& \bigl.  + 4  \sqrt{ 2 \epsilon } H M_p \left(  \Omega^\alpha{}_{m} + \frac{1}{ \left(m^{-2}\right)_{11}} \frac{\mathrm{d} 
\left(m^{-2}\right)^\alpha{}_1}{\mathrm{d}t} \right) \ R_{m \alpha m \sigma}\biggr] \nonumber  \,,
\end{align}
with $\kappa=\sqrt{2\epsilon} \Mp H/\omega_1$ the bending radius of the background trajectory in field space, and where an index $m$ means a contraction of a generic entropic index $\alpha$ with the projection vector $\left(m^{-2}\right)^\alpha{}_1 / \left(m^{-2}\right)_{11}$; for example $R_{m \sigma m \sigma}=\left(m^{-2}\right)^\alpha{}_1 \left(m^{-2}\right)^\beta{}_1/\left[\left(m^{-2}\right)_{11}\right]^2 R_{\alpha \sigma \beta \sigma}$ and there is no sum left on $m$ nor $\sigma$.
The first two lines in~\eqref{A} generalise the result of the $N_\mathrm{field}=2$ case in~\cite{Garcia-Saenz:2019njm}, and genuinely new contributions for $N_\mathrm{field} \geqslant 3$ are displayed in the third line.
Note however that even the quantities in the first two lines are affected by the whole multifield setup, for example through the projection vectors that sum over all entropic directions.
In particular, the geometrical contribution to the bispectrum is more complex than in the single-field case since not only the scalar curvature of the field space, $R_\mathrm{fs}$, enters in the final result:  Ricci scalar, contractions of the traceless Ricci tensor and of the Weyl tensor as well as their derivatives in entropic directions, are all relevant in the general case.
Eventually, note that Eqs.~\eqref{L3-EFT}-\eqref{fi} correspond to the cubic interactions of higher-derivatives, single-field theories with a $P(X,\phi)$ Lagrangian~\cite{Burrage:2011hd}, where $X$ is the standard kinetic term, $X=-1/2 g^{\mu\nu} \partial_\mu \phi \partial_\nu \phi$, provided a matching between our parameters $\left(c_s^2,A\right)$ and the derivatives of the function $P$, $\Sigma = X P_{,X} + 2 X^2 P_{,XX}$ and $\lambda=X^2 P_{,XX} +2/3 X^3 P_{,XXX}$,  which reads $c_s^2=XP_{,X}/\Sigma$ and $\left(c_s^{-2}-1 \right)A=-2\lambda / \Sigma$ (see~\cite{Garcia-Saenz:2019njm} for the simple manipulation of~\eqref{L3-EFT} needed to explicitly find the corresponding $P(X,\phi)$ cubic Lagrangian).

\paragraph{Slow-varying limit.} Until now, our results were independent of any assumption regarding the background dynamics (except that it must be such that the integrating-out procedure at leading order is justified), and in particular we have not required the slow-roll parameters to be small.
In a slow-varying approximation, the whole background dynamics is assumed to be smooth, which implies in particular that $\epsilon, \eta, s \ll 1$.
In this limit, the leading-order operators are the ones predicted by the decoupling limit of the EFToI~\cite{Creminelli:2006xe,Cheung:2007st}, \textit{i.e.} $\dot{\zeta} \left(\partial \zeta \right)^2$ and $\dot{\zeta}^3$, and the single-field effective action for $\zeta$ reads:
\begin{equation}
\mathcal{L}^{(3)}_{{\rm EFT, \, slow-varying}}= \Mp^2 \,a^3 \frac{\epsilon}{H} \left(\frac{1}{c_s^2}-1\right)  \left[ \dot{\zeta} \frac{\left(\partial\zeta\right)^2}{a^2}+ \frac{A}{c_s^2} \dot{\zeta}^3 \right] \,,
\end{equation}
where all parameters should be considered as constants for consistency.
Clearly, the interesting point of starting from the multifield setup and integrating out heavy perturbations, is that we end up with a prediction for the EFT parameters $c_s^2$ and $A$ depending on the multifield UV completion.
In the usual approach of the EFT of Inflation, those parameters are free and should be measured or constrained through cosmological observations (for the most recents constraints in the $\left(c_s^2, A\right)$-plane, see~\cite{Akrami:2019izv}).
Also, let us insist on the fact that we derived a single-field effective action for the perturbations, but that nothing forbids the homogeneous background dynamics to be genuinely multifield.
Thus, this tool is useful to compute the observable bispectrum of multifield models that have a non-trivial multifield dynamics at the level of the background, but with heavy entropic perturbations.
Lastly, if the background is not slowly-varying, or if $c_s^2$ is not much smaller than unity, one should expect all operators in~\eqref{L3-EFT} to be relevant, a situation that is not encapsulated by the decoupling limit of the EFToI.

\paragraph{Limitations of this approach.} The effective single-field theory for adiabatic perturbations during inflation that we uncovered relies on several assumptions.
More precisely, we neglected both the kinetic terms $\Box$ and the entropic mixing terms $\mathcal{O}^\Omega$ in the linear EoM of $\F^\alpha$, for all entropic perturbations.
Although this is justified when all perturbations in the entropic sector are very heavy, different regimes would be interesting to study.
First, it is possible that there is a hierarchy between the eigenstates of the mass matrix $m^2$, and that only some of the directions are heavy and can be integrated out, while the other ones should be kept dynamical.
We conjecture that the resulting theories should drastically differ depending on whether the first entropic direction, $\F^1$, which mixes with the adiabatic one, is heavy or not.
If $\F^1$ is heavy, the resulting theory should be the one of several light entropic perturbations coupled to an adiabatic perturbation that has a non-canonical kinetic term spontaneously breaking boosts symmetries, \textit{i.e.} with $c_s^2<1$.
If $\F^1$ remains light and dynamical, then $\zeta$ should keep the canonical structure with $c_s^2=1$.
A very different regime is also reached when there is a hierarchy between the kinetic terms and the entropic mixing terms, for example if the latter are larger than the former.
It would be interesting to study this other kind of ``integrating-out" procedure that incorporates  non-trivial effects coming from a fast rotation of the local entropic basis, and that were not taken into account in the present single-field effective theory analysis (although contained in the full multifield description of course).
Lastly, we assumed that both the time and the spatial derivatives (that we called ``kinetic terms") of the entropic perturbations were equally negligible.
However there is a sub-Hubble regime where gradients are much larger than time derivatives, in which case one can have the hierarchy $\frac{1}{a^3}\frac{\mathrm{d}}{\mathrm{d}t}\left(a^3 \dot{\F}^\alpha \right) \ll \left( \frac{k^2}{a^2} \delta_{\alpha\beta} + m^2_{\alpha\beta} \right) \F^\beta$ that would result in a single-field effective theory for $\zeta$ with a modified dispersion relation, \textit{i.e.} a scale-dependent speed of sound $c_s^2(k)$~\cite{Cespedes:2012hu,Gwyn:2012mw,Gong:2014rna}, as found also in genuinely single-field models at higher-order in derivatives like ghost inflation (see, \textit{e.g.}~\cite{ArkaniHamed:2003uz,Cheung:2007st,Bartolo:2010bj}).

\section{Conclusion}
\label{sec5}

In this article, we have studied the background and linear dynamics, as well as non-Gaussianities, in multifield inflation with any number $N_\mathrm{field}$ of scalar fields.
We parameterized the multifield dynamics by defining the adiabatic direction as the one instantaneously parallel to the background trajectory, and the entropic directions as the ones perpendicular (in a field-space covariant manner) to it and to one another.
Then, we identified covariant perturbations in the comoving gauge; $\zeta$ along the adiabatic direction and $\F^\alpha$ for $\alpha \in \{1,...,N_\mathrm{field}-1\}$ along the entropic ones, see Eq.~\eqref{def-F-zeta}.
Because we keep the boundary terms in the cubic action as they can play a role in the final observable bispectrum, we recalled the crucial role of the GHY boundary term in cosmology to have a well-defined variational principle. 
We computed the quadratic Lagrangian, Eqs.~\eqref{L2}-\eqref{mass-matrix}, which contains the massless adiabatic perturbation and the $N_\mathrm{field}-1$ massive entropic ones.
We found that the entropic sector mixes within itself via the non-diagonal elements of the mass matrix $m^2_{\alpha\beta}$ and the rotation of the local entropic basis $\Omega_{\alpha\beta}$, and mixes also with the adiabatic perturbation through the first entropic fluctuation $\F^1$ due to the bending of the background trajectory parameterized by $\omega_1$.
We showed in Eq.~\eqref{Ricci-decomposition2} that the geometrical contribution to the entropic mass matrix is richer than in the two-field case as not only the scalar curvature of the field space is relevant but its whole Ricci tensor should be taken into account for $N_\mathrm{field}=3$, and even its Weyl tensor for $N_\mathrm{field} \geqslant 4$.
Moreover, we explained that $\Omega$ contributes to the linear dynamics in two different ways: first squared in the effective mass matrix for large-scale fluctuations $m^2_\mathrm{eff}$, where it comes only with negative eigenvalues, but also as a linear coupling between $\F^\alpha$ and $\dot{\F}^\beta$ with $|\alpha-\beta|=1$, where its role is more complicated to analyse.
This can be seen in Eqs.~\eqref{L2}-\eqref{mass-matrix} at the level of the quadratic Lagrangian or Eq.~\eqref{eom-large-scales} at the level of the effective linear equations of motion on large scales.
Although the goal of the present article was not to address in detail the question of the linear stability of background trajectories in $N_\mathrm{field}$-inflation, we think that those findings motivate a thorough reanalysis of it.
For completeness, we also showed the quadratic Hamiltonian that may be used for in-in computations of the power spectrum in a perturbative scheme.

Next, we investigated in full generality non-Gaussianities in the form of cubic interactions, for any number of scalar fields with non-canonical kinetic terms and any potential, during inflation and without assuming particular features of the background dynamics. For this, we computed the cubic action in terms of the adiabatic perturbation and the entropic ones, a work that required many non-trivial manipulations: integrations by parts, uses of linear equations of motion. We kept the boundary terms that are needed for a careful computation of the bispectrum with the in-in formalism and identified new interactions within the entropic sector that vanish for a single isocurvature perturbation, as well as more complex geometrical interactions amongst the entropic sector and with the adiabatic perturbation. Again, we stress that the only assumption behind the general result~\eqref{L3-final}-\eqref{boundary-final} is that of small perturbations, and that otherwise it is fully generic and could be used in a large number of cases.

As a first analytical application of our general result, we explained how models that feature heavy entropic perturbations can lead to a single-field effective theory for fluctuations by integrating them out.
This procedure assumes that both the kinetic terms and the entropic mixing terms via the rotation of the entropic basis, are negligible compared to the mass matrix (that needs not be diagonal, hence entropic perturbations can mix through it).
At the quadratic level, this resulted in a single-field theory for $\zeta$ with a reduced speed of sound when $\omega_1 \neq 0$, see Eq.~\eqref{EFT-L2}.
Interestingly, the ratio of mass scales setting the deviations of $c_s^2$ from $1$ is given by $\left(m^{-2}\right)_{11} \omega_1^2$, where the inverse mass matrix $m^{-2}$ incorporates effects from the whole multifield setup, and can be large when the bending is strong. Next, after explaining that only the linear equations of motion relating the entropic fluctuations to the adiabatic perturbation were needed to compute the effectively single-field cubic interactions, we proceeded to the integrating-out procedure at that level.
As in the case of $N_\mathrm{field}=2$ scalar fields treated in our previous article~\cite{Garcia-Saenz:2019njm}, we found without further approximation a $P(X,\phi)$-like cubic Lagrangian with a direct mapping between the derivatives of $P$ and the multifield parameters, in Eq.~\eqref{L3-EFT}.
We show that in a slow-varying approximation we recover the decoupling limit of the Effective Field Theory of Inflation, but with a prediction for the otherwise unknown Wilson coefficients $\left(c_s^{-2}-1\right)$ and $A$. The latter incorporates effects from the potential as well as from the geometry of the field space, where again not only the scalar curvature matters contrary to the case $N_\mathrm{field}=2$.

We think that this work can lead to various interesting applications, both analytical and numerical.
First, we showed that the analysis of the linear stability of background trajectories has to be further questioned as soon as more than one entropic perturbation is considered.
Indeed, even the effective mass matrix on large scales may become negative due to strong turning rates of the local entropic basis, parameterized by the matrix $\Omega$ which also enters in the dynamics independently of the mass matrix.
It would be worth investigating for the first non-trivial case of $N_\mathrm{field}=3$ the effect of the torsion parameter $\omega_2$, and we conjecture about the possiblity of a ``torsion destabilisation of inflation", provided the two effects do not cancel one another.
Moreover, the geometrical contribution to the entropic mass matrix is more complex than in the case of a two-dimensional field space for which only the scalar curvature is relevant, hence it would be worth investigating in more detail its effects on the linear dynamics of perturbations and thereby possible instabilities.
We also showed the full quadratic Hamiltonian that we chose in order to define our free theory for the computation of the cubic action.
But different choices could be made in order to enable an analytical computation of the mode functions in the interaction picture: the mixing parameters (all or only some) could be considered small and the corresponding terms in the Hamiltonian treated as quadratic interactions in an in-in formulation.
With this quasi-single-field picture, one could perturbatively compute corrections to the adiabatic power spectrum due to the presence of extra scalar fields (the entropic perturbations), in the spirit of the cosmological collider program~\cite{Arkani-Hamed:2015bza}.
And since we computed the multifield cubic action in the present article, one could apply the same treatment to cubic interactions and check the effect of several extra scalar fields on the bispectrum.
More precisely, it would be extremely interesting to understand if adding several extra fields with $m\simeq H$ either strengthens the oscillatory pattern of the squeezed limit of the bispectrum~~\cite{Chen:2009we}, or completely blurs it.
For these applications, a particular regime of interest would be the one of $N_\mathrm{field} \rightarrow \infty$ for which one can hope to find universal results.
Apart from these analytical works, one could also extend the transport approach for multifield inflationary correlation functions to the comoving gauge, a numerical study that would rely on the full cubic action that we displayed in this article, and would enable to probe numerically the regime of cosmological collider physics.
Eventually, we explained in this work how to derive a single-field theory for $\zeta$ when all entropic fluctuations are heavy enough to neglect both their kinetic terms and their self-mixings.
We expect different regimes to be interesting as well, for example if the mixings are not negligible one would have to define a different integrating out procedure.
Moreover, it could happen that only some of the entropic fluctuations are heavy and others are light, in which case a partial integration of entropic modes could be relevant.
We leave these exciting prospects for future works.

\acknowledgments

I would like to thank Sebastian Garcia-Saenz and S\'ebastien Renaux-Petel for initiating this project in the case of $N_\mathrm{field}=2$ and always providing me with encouragements for the present generalisation to any number $N_\mathrm{field}$ of scalars.
I also thank them together with 
Jacopo Fumagalli,
Thomas Hussenot,
Fran\c{c}ois Larrouturou,
and Lukas Witkowski
%S\' ebastien Renaux-Petel and Sebastian Garcia-Saenz
for useful comments on the present manuscript and its content,
as well as Gonzalo Palma for interesting discussions about the early stage of this project when he kindly invited me to visit him in Santiago in January 2020.
I am supported by the European Research Council under the European Union's Horizon 2020 research and innovation programme (grant agreement No 758792, project GEODESI).

%\subsection{Special cases $N=2,3$ and relation to previous works}

%\subsubsection{$N=2$}

%paper with Seb$\&$Seb / cosmological collider physics: Wang$\&$Chen, the 3 Japanese, Dong-Gang

%\subsubsection{$N=3$}

%Palma$\&$Cespedes: second order action and eom, single-field EFT / cosmological collider physics

%\subsection{Squeezed limit of the $\zeta$ bispectrum: general case}

\bibliographystyle{JHEP}
\bibliography{main}

\end{document}